\documentclass[preprint,12pt]{elsarticle}

\pdfoutput=1
\usepackage{lineno}
\usepackage{subcaption}
\usepackage{hyperref}
\hypersetup{pdfstartview=}

\usepackage{color}
\usepackage{amssymb,amsmath}

\modulolinenumbers[10]

\journal{}

\begin{document}

\begin{frontmatter}

\title{Bulk and Surface Event Identification 
in p-type Germanium Detectors}

%% or include affiliations in footnotes:
\author[address0,address1]{L.T.~Yang}
\author[address2]{H.B.~Li\corref{correspondingauthor}}
\ead{lihb@gate.sinica.edu.tw}
\author[address2]{H.T.~Wong}
\author[address2,address3]{ M.~Agartioglu }
\author[address2]{J.H.~Chen}
\author[address1]{L.P.~Jia}
\author[address1]{H.~Jiang}
\author[address1]{J.~Li}
\author[address2]{F.K.~Lin}
\author[address4]{S.T.~Lin}
\author[address1,address4]{S.K.~Liu}
\author[address0]{J.L.~Ma}
\author[address2,address3]{B.~Sevda}
\author[address5]{V.~Sharma}
\author[address5]{L.~Singh}
\author[address2,address5]{M.K.~Singh}
\author[address2,address5]{M.K.~Singh}
\author[address2,address5]{A.K.~Soma}
\author[address2,address3]{A.~Sonay}
\author[address2]{S.W.~Yang}
\author[address1]{L.~Wang}
\author[address0,address1,address6,address7]{Q.~Wang}
\author[address1]{Q.~Yue\corref{correspondingauthor}}
\ead{yueq@mail.tsinghua.edu.cn}
\author[address1]{W.~Zhao}

\cortext[correspondingauthor]{Corresponding author}

\address[address0]{Department of Physics,
Tsinghua University, Beijing 100084}
\address[address1]{Key Laboratory of Particle and Radiation 
Imaging (Ministry of Education) and Department of 
Engineering Physics, Tsinghua University, Beijing 100084}
\address[address2]{Institute of Physics, Academia Sinica,
Taipei 11529}
\address[address3]{Department of Physics,
Dokuz Eyl\"{u}l University, Buca, \.{I}zmir 35160}
\address[address4]{College of Physical Science and Technology, 
Sichuan University, Chengdu 610064}
\address[address5]{Department of Physics, 
Banaras Hindu University, Varanasi 221005}
\address[address6]{Center for High Energy Physics,
Tsinghua University, Beijing 100084}
\address[address7]{Collaborative Innovation Center of Quantum Matter,
Beijing 100084}

\begin{abstract}

The p-type point-contact germanium detectors have been adopted for light
dark matter WIMP searches and the studies of low energy neutrino physics.
These detectors exhibit anomalous behavior to events located 
at the surface layer.
The previous spectral shape method to identify these surface events
from the bulk signals relies on spectral shape assumptions 
and the use of external calibration sources.
We report an improved method in separating them 
by taking the ratios among different categories of 
{\it in situ} event samples as calibration sources.	
Data from CDEX-1 and TEXONO experiments	are re-examined	using	
the ratio method. Results are shown to be consistent with 
the spectral shape method.

\end{abstract}

\begin{keyword}
Dark matter\sep Radiation detector \sep Pulse Shape Analysis
%%\MSC[2010] 00-01\sep  99-00
\end{keyword}

\end{frontmatter}

%%\newpage

\linenumbers

\section{Introduction}\label{section:introduction}

The p-type point-contact germanium detectors 
({\it p}Ge)~\cite{Luke:1989,Barbeau:2007} possess the
merits of low intrinsic radioactivity background 
and excellent energy threshold in the sub-keV energy range. 
They have been used in rare-event detection
experiments, such as the search of “light” 
Weakly Interacting Massive Particles (WIMPs) 
with mass range 1~GeV$<m_{\chi}<$10~GeV, 
searches of solar and dark matter axions~\cite{Liu:2017}, 
as well as studies of
neutrino electromagnetic properties 
and neutrino-nucleus coherent
scattering with reactor neutrinos~\cite{Yue:2004,Wong:2006,Soma:2016}.  

Anomalous excess events from the CoGeNT experiment with 
{\it p}Ge~\cite{Aalseth:2011a,Aalseth:2013,Aalseth:2014}
have be taken as signatures of light WIMPs. This interpretation is
contradicted by the CDEX-1 experiment at China Jinping Underground 
Laboratory~\cite{Zhao:2013,Yue:2014a,Zhao:2016} 
and the TEXONO experiment at the Kuo-Sheng Reactor
Neutrino Laboratory~\cite{Li:2013a,Li:2014a}, 
also using {\it p}Ge as target.

Central to the discussion is 
the treatment of anomalous behavior of surface events in 
{\it p}Ge~\cite{Li:2014a,Martin:2012,Aguayo:2013,Aalseth:2015a},
incorrect or incomplete correction of these effects may lead 
to false interpretation of the data and limit the experimental sensitivities. 
The analysis of anomalous surface events 
and the differentiation between bulk and surface events (BSD) in {\it p}Ge is 
therefore crucial to realize the full potentials of this novel detector technique.

The anomalous surface events were studied 
with the ``spectral shape method'' in an
early work~\cite{Li:2014a}. 
However, there are several inadequacies with this approach.
In this article, we report an improved 
``ratio method''  to address these deficiencies,
in which {\it in situ} data provide additional important 
constraints and information. 

The article is organized as follows.
The physics of anomalous surface events in {\it{p}}Ge detectors 
is described in Section~\ref{section:BS}. 
The features of uniformity of measured rise-time distributions 
among different event samples are discussed in Section~\ref{section::risetime}.
The spectral shape method is summarized in Section~\ref{section:old_BS}, 
followed by detailed discussions on the ratio method in 
Section~\ref{section:new_BS_method}. The application to the published 
data and comparison of their results are discussed in 
Section~\ref{section:on_c1a_data}.

We follow the notations of 
earlier work~\cite{Yue:2014a,Zhao:2016,Li:2014a}, where
$\rm{AC}$ and $\rm{CR}$ denote
the anti-Compton detector and the cosmic-ray veto systems, respectively,
while the superscript $\rm{-(+)}$ corresponds to 
anti-coincidence (coincidence) with the {\it{p}}Ge signals. 
Neutrino- and WIMP-induced candidate events would therefore
manifest as $\rm{AC}^{-}$ 
and $\rm{CR}^{-}{\otimes}\rm{AC}^{-}$ 
in the CDEX-1 and TEXONO data, respectively.

\begin{figure}[!htbp]
\centering\includegraphics[width=0.8\linewidth]{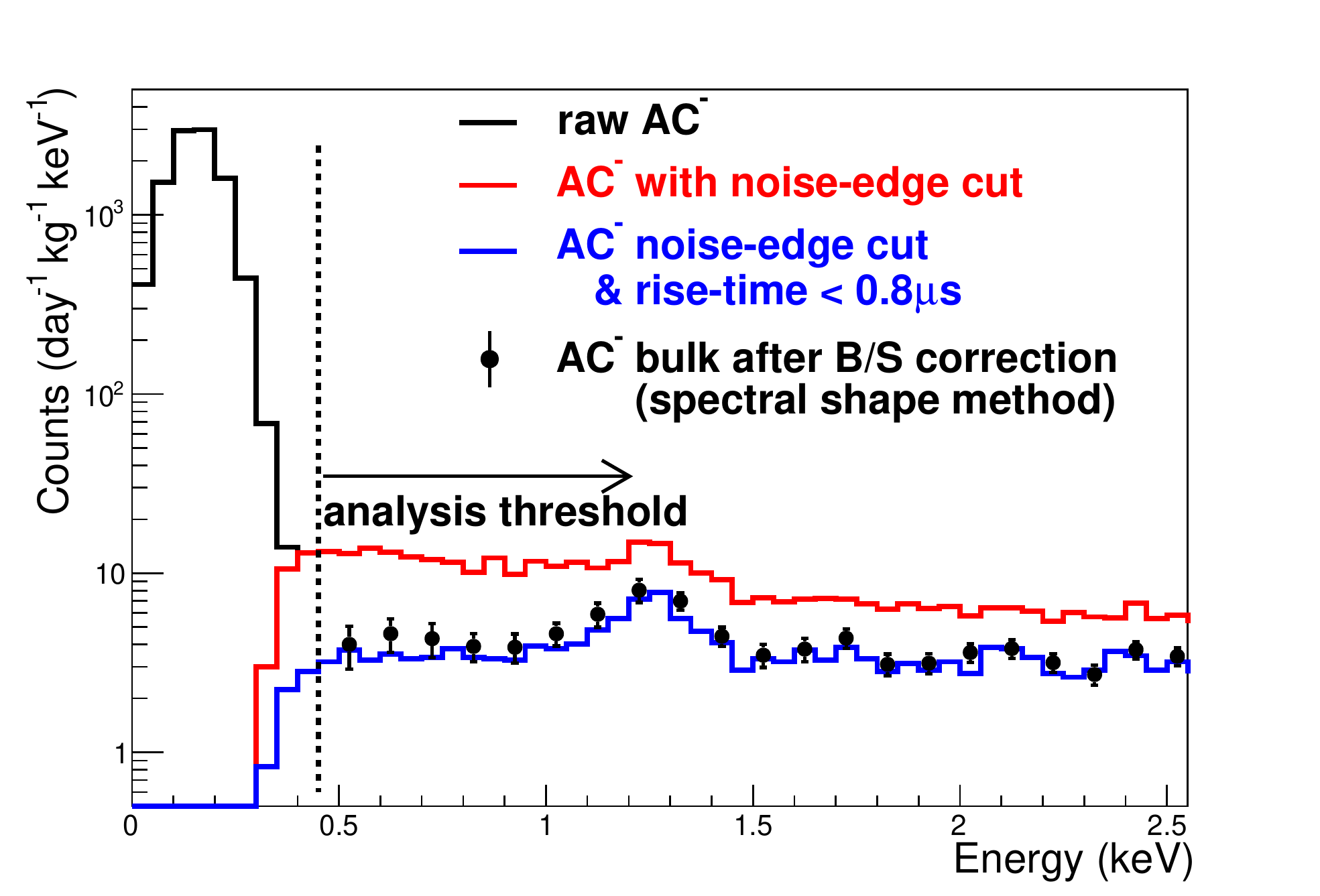}
\caption{
AC$^{-}$ spectra of CDEX-1 experiment at various stages~\cite{Zhao:2016}:
before performing noise-edge cut, after noise-edge cut,
spectrum after noise-edge cut and event selection by
$\tau$$<$0.7$\mu{s}$, spectrum after noise-edge cut and B/S correction
of spectral shape method (described in Section~\ref{section:old_BS})
It shown that the noise-edge is around 350~eV, the analysis
threshold in this article is set at 450~eV.
}
\label{fig:c1a_noise_edge}
\end{figure}

$\rm{AC}^{-}$ spectra of CDEX-1 at various stages of event selection
are shown in Figure~\ref{fig:c1a_noise_edge}.

\section{Anomalous Surface Events in {\it{p}}Ge Detectors}
\label{section:BS}

The anomalous surface charge collection effect
in {\it p}Ge was noted 
in early literature~\cite{Soma:2016}.
Recent interest of adopting
the {\it p}Ge techniques in dark matter experiments gives rise to 
thorough studies~\cite{Li:2014a,Martin:2012,Aguayo:2013}.  

The $n^+$ surface electrodes of {\it p}Ge
are fabricated by lithium diffusion and have
a typical thickness of $\sim$1~mm~\cite{Martin:2012,Jiang:2016}.
Electron-hole pairs produced at the
surface (S) layer in {\it p}Ge
are subjected to a weaker drift field
than those in the bulk volume (B).
A fraction of the pairs will recombine while the
residuals will induce signals which are weaker
and slower than those originated in B.
The S-events would therefore exhibit slower rise-time and
partial charge collection compared to B-events.
The charge collection efficiency as a function of the depth 
of the surface was recently measured and simulated~\cite{Ma:2017}. 
The n-type point-contact germanium detectors, 
having micron-sized $p^+$ surface electrode due to boron-implantation,
do not exhibit anomalous surface events~\cite{Soma:2016}.

\begin{figure}
\centering
\begin{subfigure}{0.5\textwidth}
  \centering
  \includegraphics[width=1.08\linewidth]{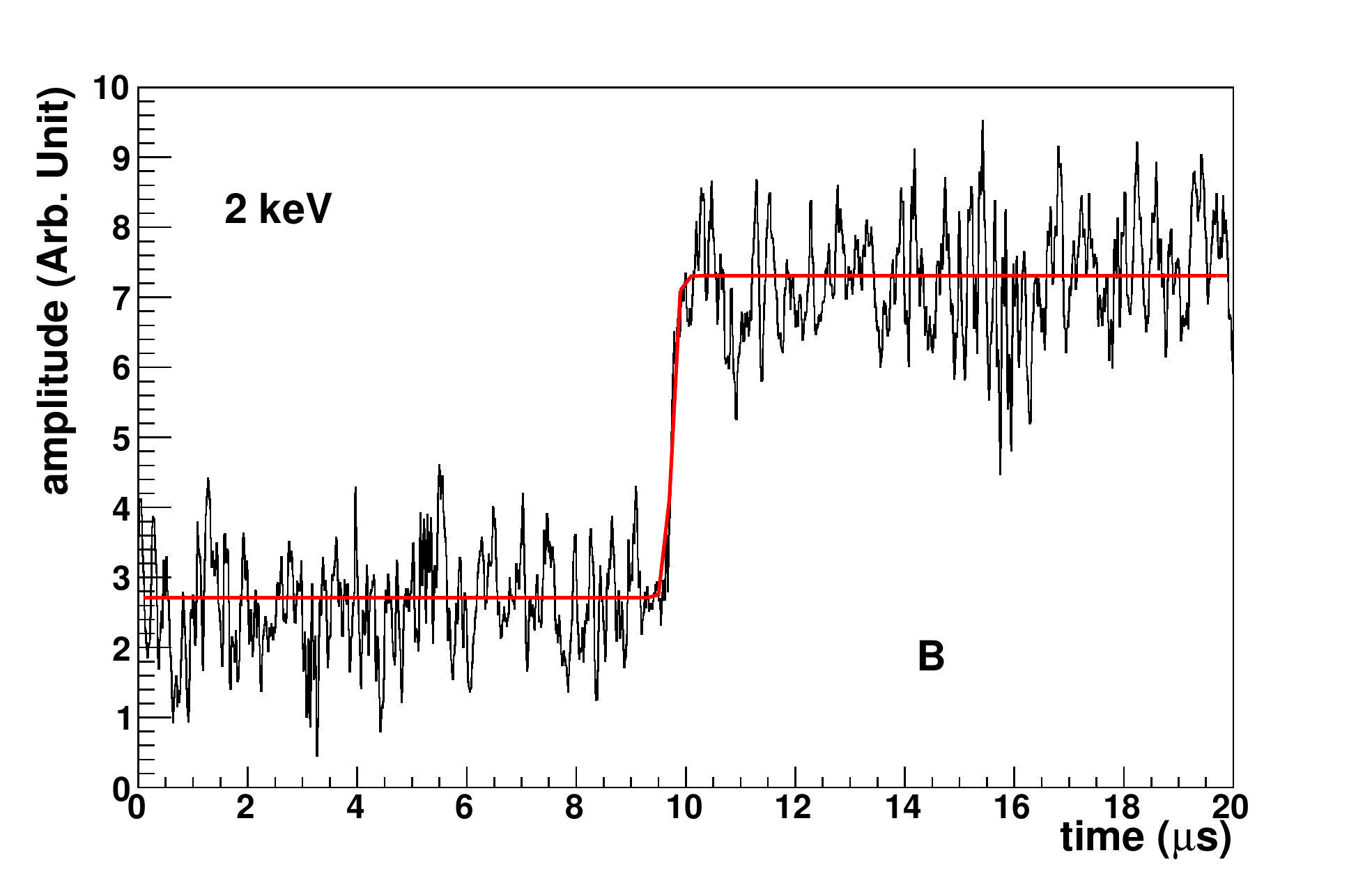}
  \caption{}
\end{subfigure}%
\begin{subfigure}{0.5\textwidth}
  \centering
  \includegraphics[width=1.08\linewidth]{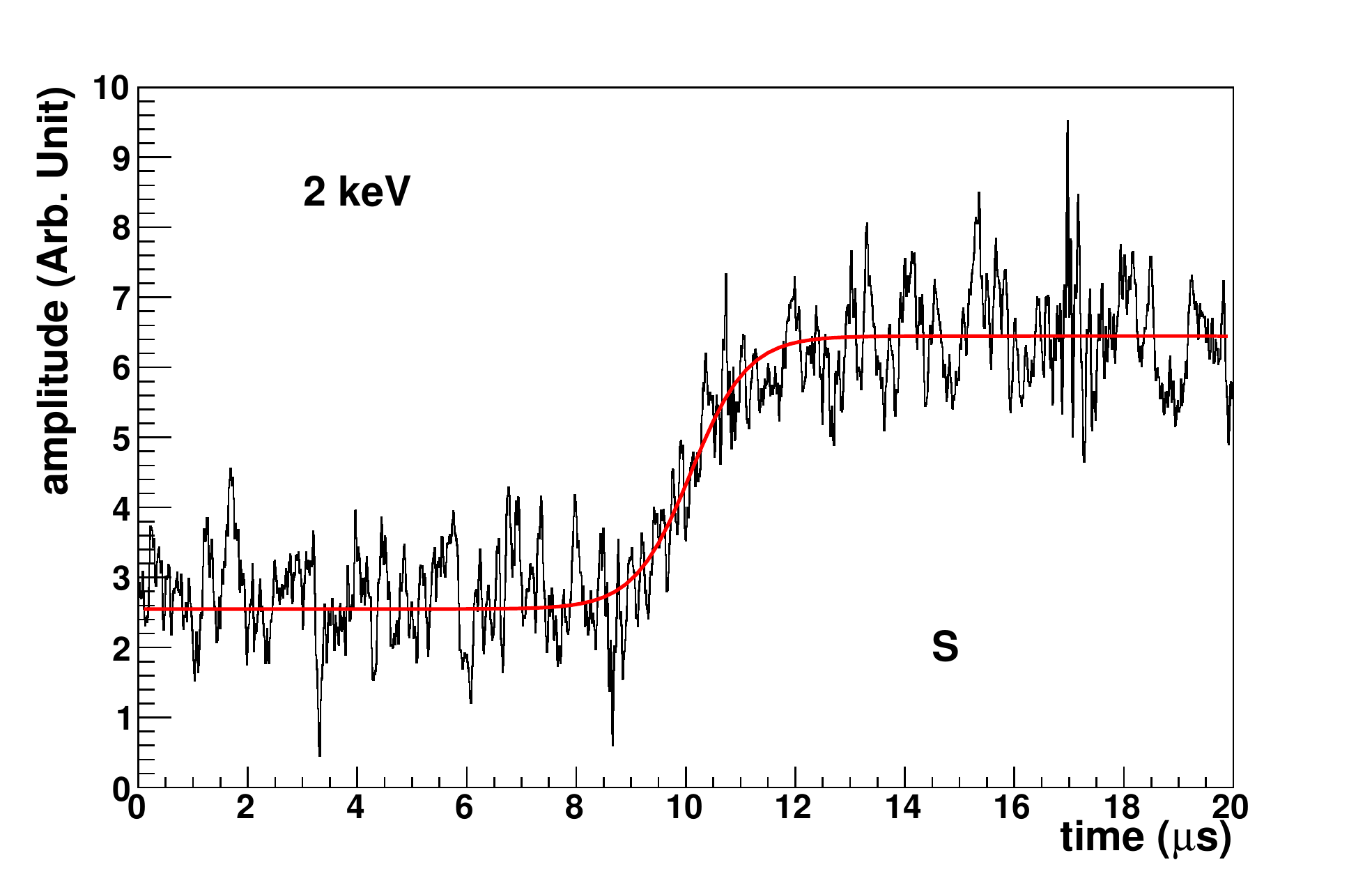}
  \caption{}
\end{subfigure}
\centering
\begin{subfigure}{0.5\textwidth}
  \centering
  \includegraphics[width=1.08\linewidth]{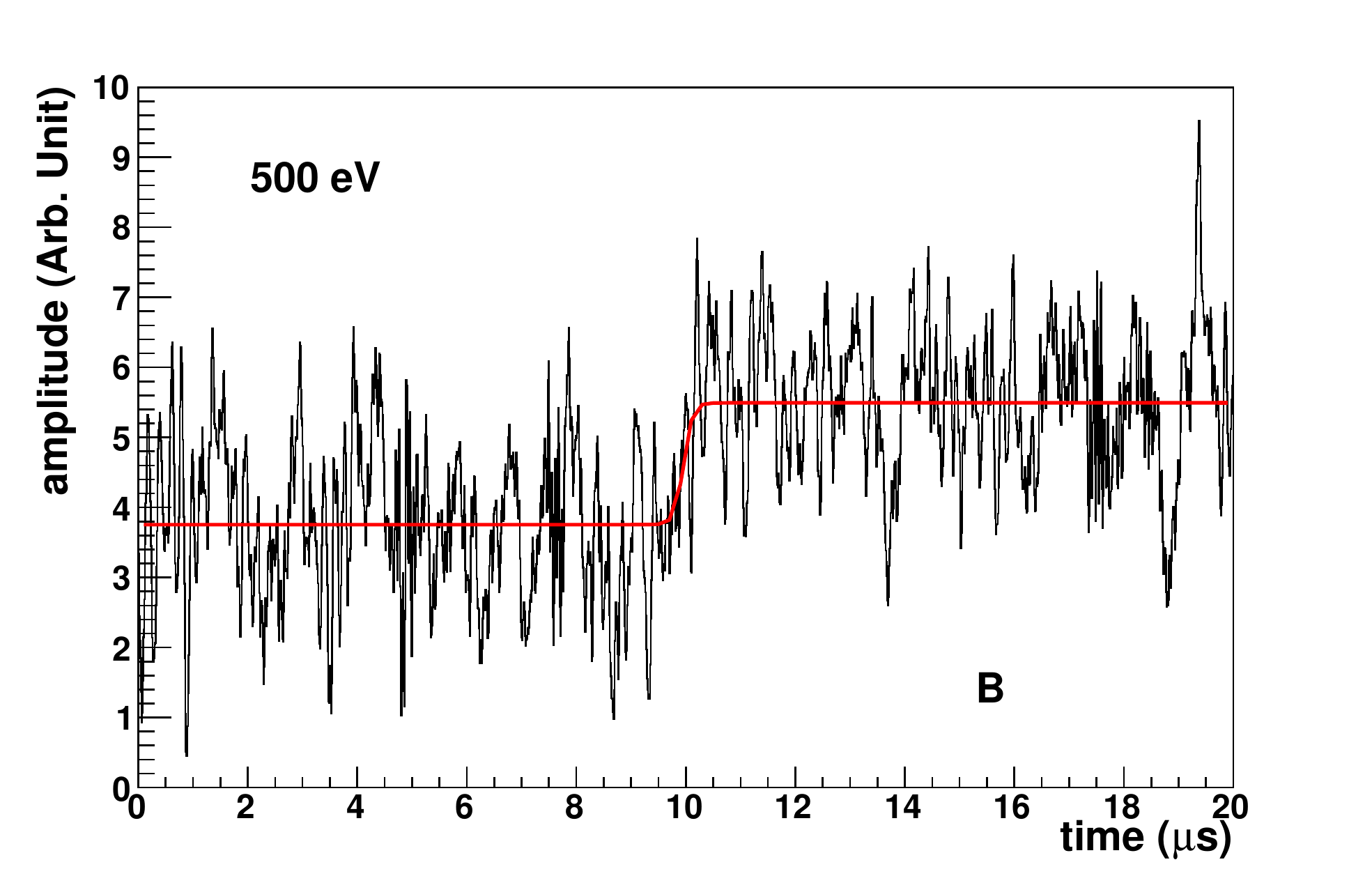}
  \caption{}
\end{subfigure}%
%%\hspace*{\fill}
\begin{subfigure}{0.5\textwidth}
  \centering
  \includegraphics[width=1.08\linewidth]{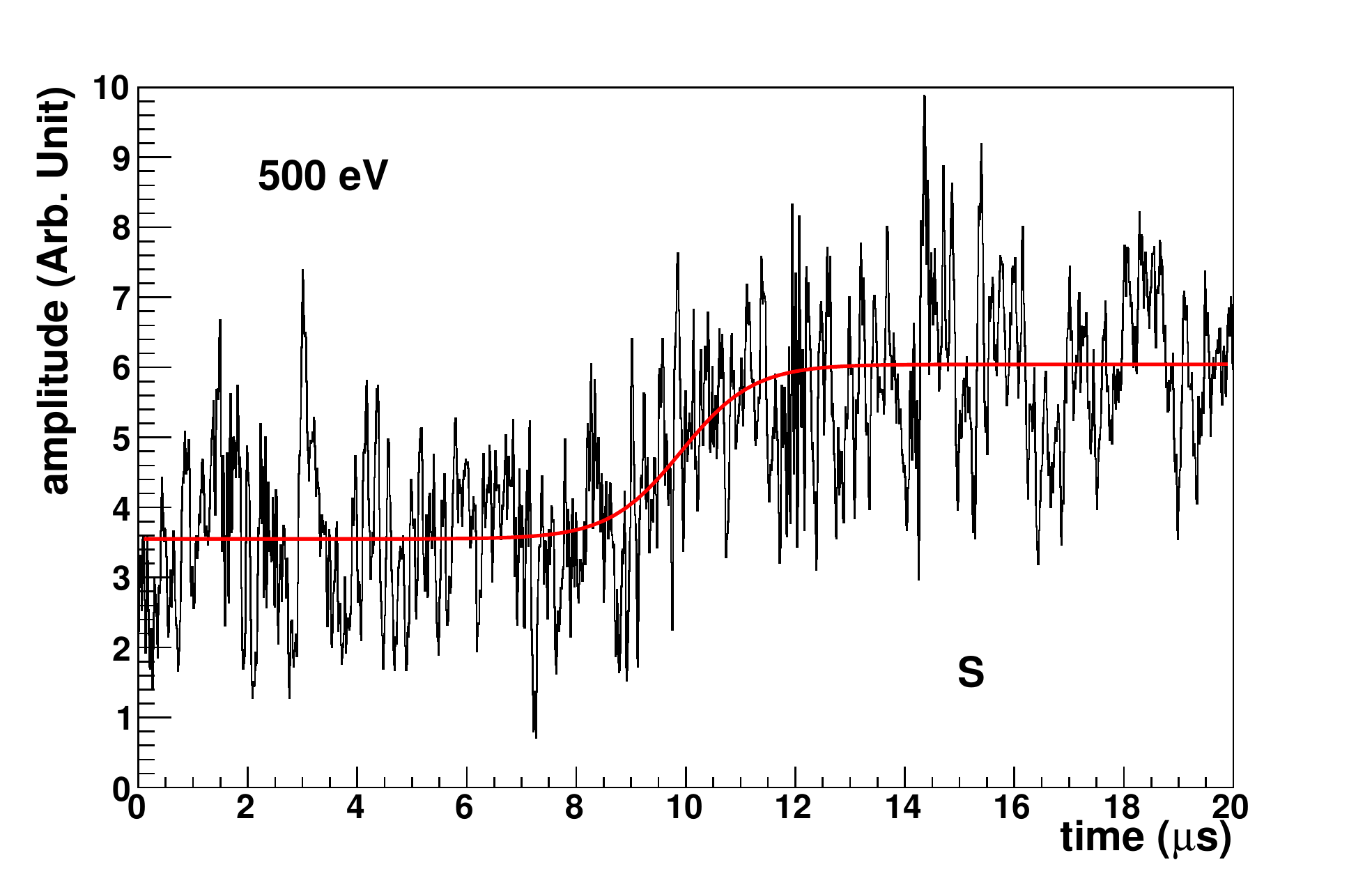}
  \caption{}
\end{subfigure}
\caption{Amplitude versus time 
of typical {\it p}Ge signals from CDEX-1 with
(a) fast rise-time, from a 2~keV bulk event,
(b) slow rise-time, from a 2~keV surface event,
(c) fast rise-time, from a 0.5~keV bulk event
and
(d) slow rise-time, from a 0.5~keV surface event.
The best-fit profiles from Eq.~\ref{eq::taufct}
are superimposed.
}
\label{fig:fast_slow_pulse_example}
\end{figure}

\begin{figure}[!htbp]
\centering\includegraphics[width=0.8\linewidth]{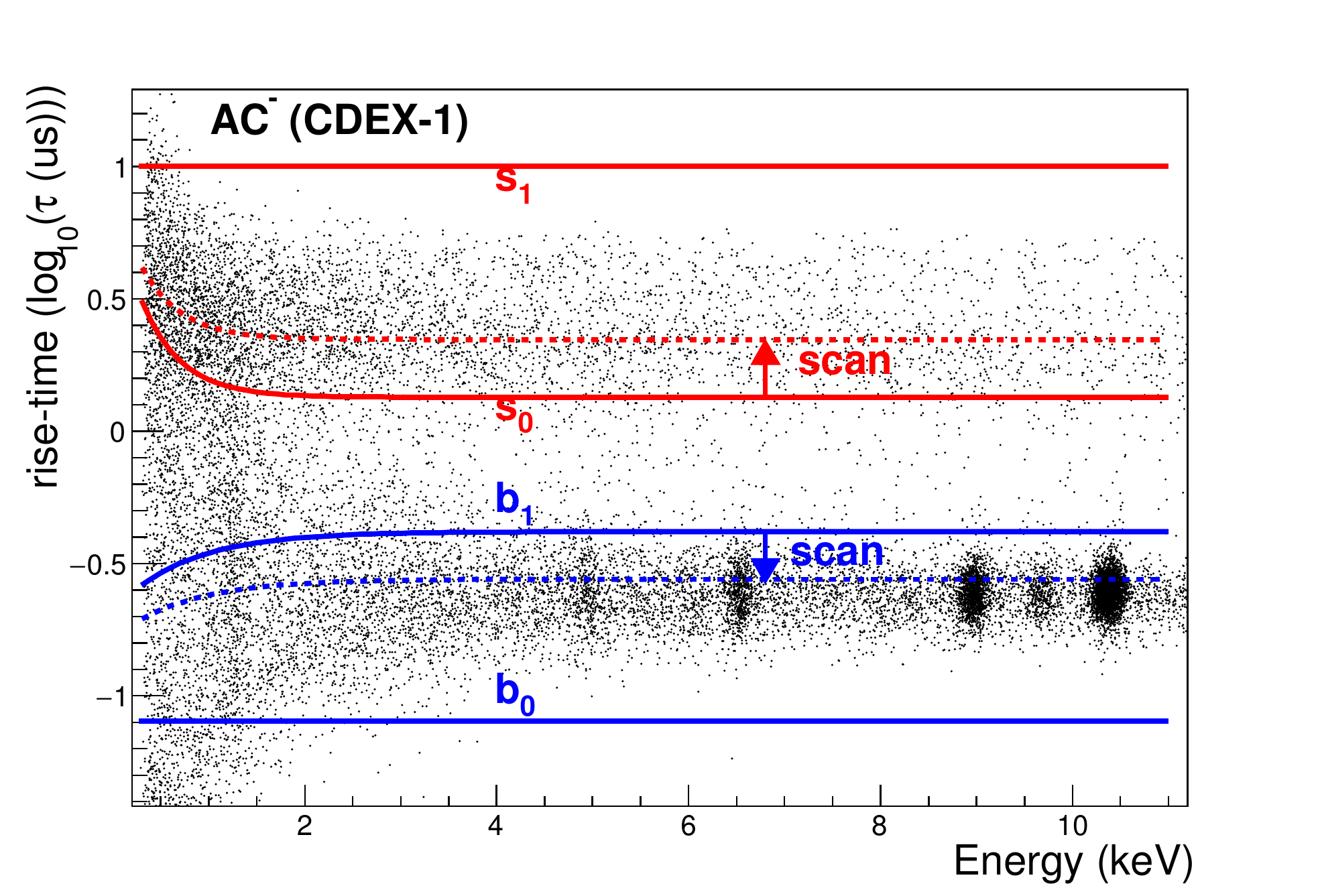}
\caption{
Rise-time versus energy scatter plot for the WIMP-induced candidate
events based on $\rm{AC}^{-}$ selection in CDEX-1 data.
The lines label
$b_{0}$, $b_{1}$, $s_{0}$, $s_{1}$
are described in Section~\ref{subsection:evaluation}
and Section~\ref{subsection::uncertainties} for systematic checks.
}
\label{fig:rt_vs_lowE}
\end{figure}

Electronic signals are induced by the drifting charges.
The signal rise-time ($\tau$) can be parametrized by
the hyperbolic tangent function
\begin{equation}
\frac{1}{2}~{A_0}
\times ~ \tanh(\frac{t-{t_0}}{\tau}) ~ + ~ {P_0} ~~ ,
\label{eq::taufct}
\end{equation}
where ${A_0}$, ${P_0}$ and ${t_0}$ are
the amplitude, pedestal offset and timing offset, respectively.
Typical examples of B- and S-events, showing both
their raw pulses and the fitted-profiles, at 2~keV and 0.5~keV,
are illustrated in 
Figures~\ref{fig:fast_slow_pulse_example}a,b,c\&d, respectively.
A typical rise-time versus energy scatter plot is shown in Figure~\ref{fig:rt_vs_lowE}.

At high energy where S/N$>>$1,
the fits are in excellent agreement with data
indicating that Eq.~\ref{eq::taufct}
is an appropriate description of the rise-time
of physics events.
However, at low energy ($<$2~keV) where the
signal amplitude is comparable to that of electronic pedestal noise, 
the B- and S-events could be falsely identified,
giving rise to cross-contaminations.
Software algorithms have to be applied to account for and
correct these effects.

Typical pulses at energy near threshold are 
depicted in Figure~\ref{fig:fast_slow_pulse_example}c\&d.
The analysis threshold of 
450~eV is well above the RMS of pedestal noise of 
62~eV and
measureable noise-edge of 350~eV, as shown in Figure~\ref{fig:c1a_noise_edge}.
Assuming one exponentially decreasing noise contribution
the fraction of noise events is $<$1\% at 450~eV.

\section{Rise-Time Uniformity}
\label{section::risetime}

The validity of the software algorithms discussed in this 
article to differentiate bulk and surface events stands 
on the uniformity of $\tau$-distributions among both 
electronic and nuclear recoil events in describing the 
data to the desired level of accuracy. 

%modify v22 : new
Events produced by different particles (electrons, gammas, neutrons)
exhibit similar bulk rise-time distributions in Ge detectors with the current
generation of technology. Previous work indicated no difference
of bulk rise-time distributions for $\gamma$-sources and nuclear recoil~\cite{Baudis:1998},
and recent work reported that electron and nuclear                      
events may differ in their rise-time by about $\sim$10~ns due to plasma 
effects~\cite{Wei:2016}, 
much faster than the typical Ge detectors rise-time of $\sim$1~${\mu}s$.
Differentiation of these signals are at the forefront of research, 
the success of which
would represents a major advance in Ge-detector techniques and
applications. 

%modify v22 : new
Bulk electron and nuclear recoil events are therefore
indistinguishable from their rise-time distributions in Ge ionization detector~\cite{Baudis:1998}.
Accordingly, rise-time distributions are the same at different B-regions 
while different depth in
S-layers give different rise-time distributions due to 
the difference in diffusion time of electrons in the surface-inactive 
regions to the bulk-drifting volume~\cite{Soma:2016,Zhao:2016,Li:2014a}.
The consequences of both are 
that the rise-time distributions are:
(a) uniform for B-events for all sources while 
(b) different for S-events due to different event-depth distributions for 
sources of different energy.

%%MODIFY v24
Non-uniformity of surface rise-time distributions is corrected by calibration sources selection,
as discussed in details in Section~\ref{section:uniformity_and_calibration}.
The selection is data/experiment dependent, not universally applicable to all analysis.

The understanding of nature of rise-time distributions is
beyond the scope in this analysis.
An {\it{ab initio}} approach by simulation of behavior of particles in {\it p}Ge and
configuration of {\it p}Ge would provide an alternative way to understand and address
the B/S issue, though the current accuracies do not match the data-driven approaches 
discussed
%%ADD v24
in this article.

\section{Bulk-Surface Differentiation: Spectral Shape Method}
\label{section:old_BS}

The spectral shape method is 
a cut-based algorithm~\cite{Li:2014a} developed
to perform BSD
for light WIMP searches with 
the CDEX-1~\cite{Yue:2014a,Zhao:2016} 
and TEXONO~\cite{Li:2013a} data.

Two parameters have to be derived:
the B-signal retaining and 
S-background suppression
efficiencies, denoted by $\epsilon_{BS}$ and 
$\lambda_{BS}$, respectively.
The efficiency-corrected ``real'' B- and S-rates
$( B_r , S_r )$ are related to measured rates $( B_m , S_m )$ via:
\begin{align}
B_{m}=\epsilon_{BS}B_{r}+(1-\lambda_{BS})S_{r} \nonumber \\
S_{m}=\lambda_{BS}S_{r}+(1-\epsilon_{BS})B_{r}, \label{eq:bm_to_br}
\end{align}
with an additional unitary constrain of $B_{m}+S_{m}=B_{r}+S_{r}$.

The solutions of Eq.~\ref{eq:bm_to_br} are:
\begin{align} 
B_{r}=\frac{\lambda_{BS}B_{m}-(1-\lambda_{BS})S_{m}}
            {\epsilon_{BS}+\lambda_{BS}-1} \nonumber \\
S_{r}=\frac{\epsilon_{BS}S_{m}-(1-\epsilon_{BS})B_{m}}
            {\epsilon_{BS}+\lambda_{BS}-1}. \label{eq:bm_to_br_solutions}
\end{align}

Two components contribute to $B_r$($S_r$).
The first positive term accounts for 
the loss of efficiency in the measurement of $B_m$($S_m$),
while the second negative term corrects  
misidentification due to contamination effects.
Both ($\epsilon_{BS} , \lambda_{BS}$) factors 
should be properly accounted for in
order to provide correct measurements of the 
energy spectra for bulk events.

In order to solve Eq.~\ref{eq:bm_to_br}
for the two unknown parameters ($\epsilon_{BS} , \lambda_{BS}$), 
at least two sources with different but known
B- to S-event ratio are required.
Four calibration sources 
($^{137}\rm{Cs}$, $^{241}\rm{Am}$, $^{57}\rm{Co}$ 
and $^{60}\rm{Co}$)~\cite{Yue:2014a,Zhao:2016} were used
in CDEX-1 analysis. 
The $B_{r}$ spectra of these sources were evaluated 
by full GEANT4 simulation,
so that ($\epsilon_{BS}$, $\lambda_{BS}$)
were derived having the corresponding measured $B_m$. 
The WIMP candidate data and ambient gamma background were then 
corrected by $\epsilon_{BS}$ and $\lambda_{BS}$.

However, there are several deficiencies 
with the spectral shape method:
\begin{enumerate} [(1)]
\item \textbf{Spectral shape assumption.} \\
Only sources with known spectral shape from simulations could be 
used as calibration data. 
{\it In situ} data like ambient background
from $\gamma$-radioactivity do not contribute to calibration.
This poses potential problems in long term data taking, such 
that data with external calibration source have to be taken 
at regular intervals and stability has to be assumed in
between them.

The $B_{r}$-spectra of calibration sources are evaluated 
from GEANT4 simulation, 
which depends on the detector structure and physics 
process subroutines adopted.
In realistic data taking, there are additional contributions 
to $B_{m}$ due to cosmic-induced or ambient background 
which would introduce new error sources. 

\item \textbf{Normalization assumption.}\\
The spectra of $B_m$ and $B_r$ in calibration have to be normalized.
The chosen scheme is to assume $\epsilon_{BS}$ and $\lambda_{BS}$ 
to be 1~\cite{Zhao:2016,Li:2014a} (that is, perfect differentiation)
at the high energy range of
$\sim$2$-$4~keV for the various calibration data.
This assumption, while reasonable, may introduce 
additional uncertainties.

\item \textbf{Singularity problem.}\\
The solutions of Eq.~\ref{eq:bm_to_br} are undefined 
and the uncertainties becomes infinite 
when $\epsilon_{BS}+\lambda_{BS}$ approaches 1. 

\end{enumerate}

To address these drawbacks in performing BSD,
we develop the ratio method 
to be discussed in the following sections.

\section{Bulk-Surface Discrimination: Ratio Method}
\label{section:new_BS_method}

\subsection{Concept and Formulation}\label{subsection:best_fit}

Adopted data samples include calibration and {\it in situ} 
physics events, and are represented by index $i$. The goal
of the analysis is to extract information on
the B- and S-event distributions 
which are in general functions of $( E , \tau )$ 
and denoted as
$N_{B i}(E,\tau)$ and $N_{S i}(E,\tau)$, respectively.
The relevant quantities for physics analysis are 
the real B- and S-rates which corresponds to, respectively,
\begin{equation}
B_{r i}(E)=\int_{all~\tau}N_{B i}(E,\tau) d \tau 
~~~~ {\rm and } ~~~~~
S_{r i}(E)=\int_{all~\tau}N_{S i}(E,\tau) d \tau  ~~ .
  \label{eq:Br_Sr}
\end{equation}
In particular, $B_{r i} (E)$ would be 
the neutrino- and WIMP-induced candidate spectra
where $i$ corresponds to the data sample 
surviving the electronic noise, 
cosmic-ray and anti-Compton veto selections.

One can write
\begin{align}
N_{B i}(E,\tau)=\beta_{i}(E)f_{B}(E,\tau) \nonumber \\
N_{S i}(E,\tau)=\xi_{i}(E)f_{S}(E,\tau). 
\label{eq:common_b_plus_s_0}
\end{align}
where  $\beta_{i} (E)$ and $\xi_{i} (E)$ 
are $\tau$-independent scaling factors
proportional to the B- and S-event rates.
Evidence for independence of the rise-time distributions 
$f_{B}(E,\tau)$ and $f_{S}(E,\tau)$
from different particle interactions is discussed in 
Section~\ref{section::risetime}.

The measured count rate of the $i^{th}$-sources as
functions of $E$ and $\tau$ is therefore
\begin{eqnarray} 
N_{i}(E,\tau)& = & N_{B i}(E,\tau)+N_{S i}(E,\tau) \nonumber \\
& = & \beta_{i}(E)f_{B}(E,\tau)+\xi_{i}(E)f_{S}(E,\tau) ~~ .
\label{eq:common_b_plus_s}
\end{eqnarray}

To obtain the desired output 
of $N_{B i}(E,\tau)$ and $N_{S i}(E,\tau)$,
additional constraints must be provided
to Eq.~\ref{eq:common_b_plus_s}.  
For instance, modeling assumptions were made to
$f_{B}(E,\tau)$ and $f_{S}(E,\tau)$ 
in the CoGeNT experiment~\cite{Aalseth:2015a},
while the spectral shape method adopted in the 
TEXONO~\cite{Li:2014a} and 
CDEX-1~\cite{Yue:2014a,Zhao:2016} analysis
stands on having $\int{N_{B i}(E,\tau)}d\tau$ values
known by simulations for certain calibration sources.

For a collection of different sources with differing Bulk to 
Surface event ratios Eq.~\ref{eq:common_b_plus_s}
can be used to find $N_{B i}$ and $N_{S i}$ by $\chi^{2}$ 
minimization of the right
hand-side of the equation, i. e.,
$\beta_{i}(E)f_{B}(E,\tau)+\xi_{i}(E)f_{S}(E,\tau)$
\begin{equation} \label{eq:chi2_fB_fS}
\chi^{2} ( E , \tau ) = \sum_{i} 
\frac{ [ \beta_{i} (E) f_{B} ( E , \tau )
+\xi_{i} (E) f_{S} ( E , \tau ) - N_{i} ( E , \tau )]^{2} }
{ \Delta{N}_{i} ( E , \tau ) ^2 }  ~~ .
\end{equation}

The absolute values of $\beta_{i} (E)$ and $\xi_{i} (E)$
are not relevant to this analysis. The important values are 
$\beta_{i}(E)f_{B}(E,\tau)$ and $\xi_{i}(E)f_{S}(E,\tau)$.

In fact, we are free to choose $\beta_{i} (E)$ and $\xi_{i} (E)$,
as long as they satisfy Eq.~\ref{eq:common_b_plus_s_0},
which is equivalent to
\begin{align}
\frac{N_{B i}(E,\tau)}{N_{B j}(E,\tau)} =
\frac{\beta_{i}(E)}{\beta_{j}(E)} \nonumber \\
\frac{N_{S i}(E,\tau)}{N_{S j}(E,\tau)} =
\frac{\xi_{i}(E)}{\xi_{j}(E)} .
\label{eq:ratio_scaling_0}
\end{align}
This $\tau$-independent ratios are the basis of the ratio method.

\begin{figure}
\centering
\begin{subfigure}{0.5\textwidth}
  \centering
  \includegraphics[width=1.08\linewidth]{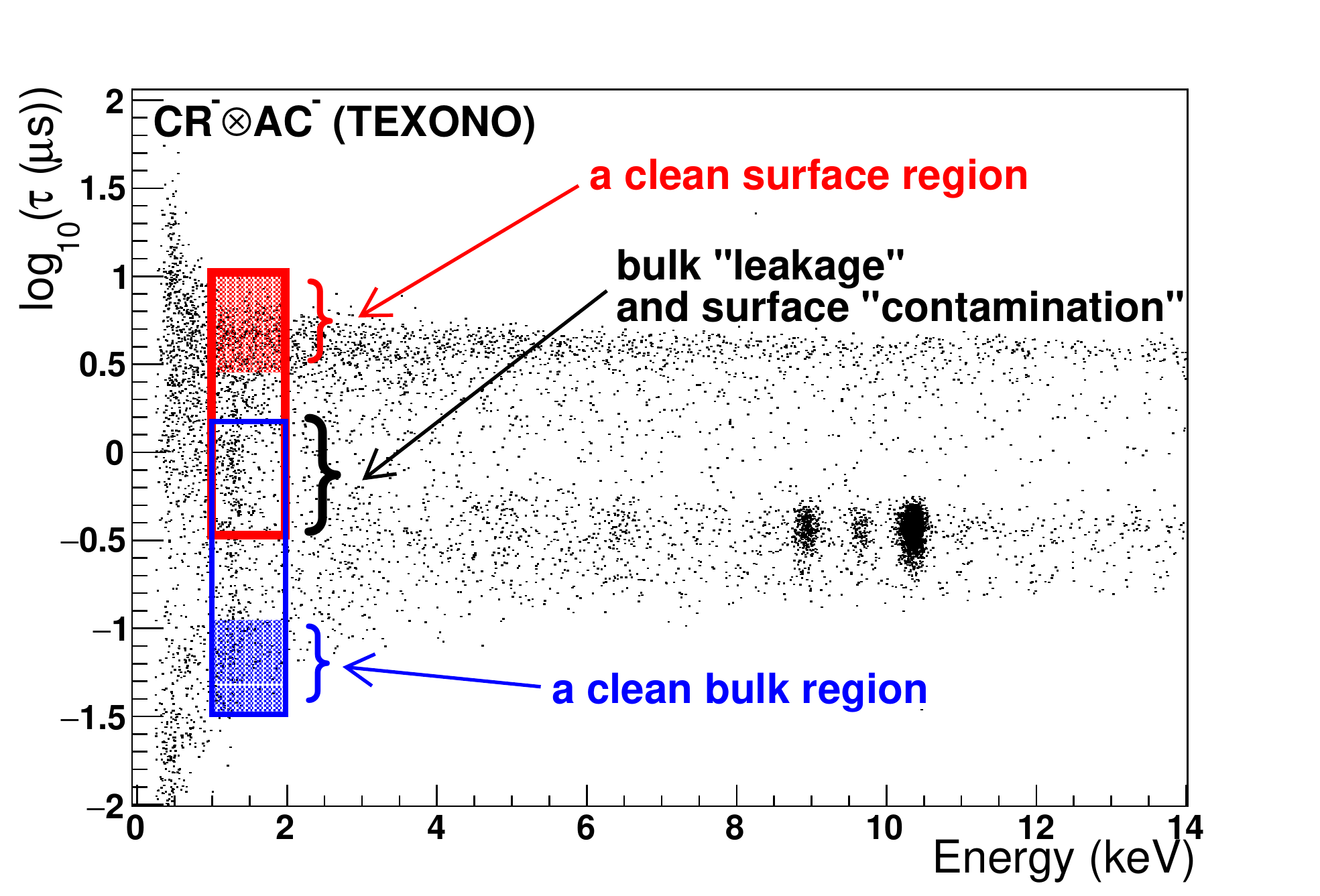}
  \caption{}
\end{subfigure}%
\begin{subfigure}{0.5\textwidth}
  \centering
  \includegraphics[width=1.08\linewidth]{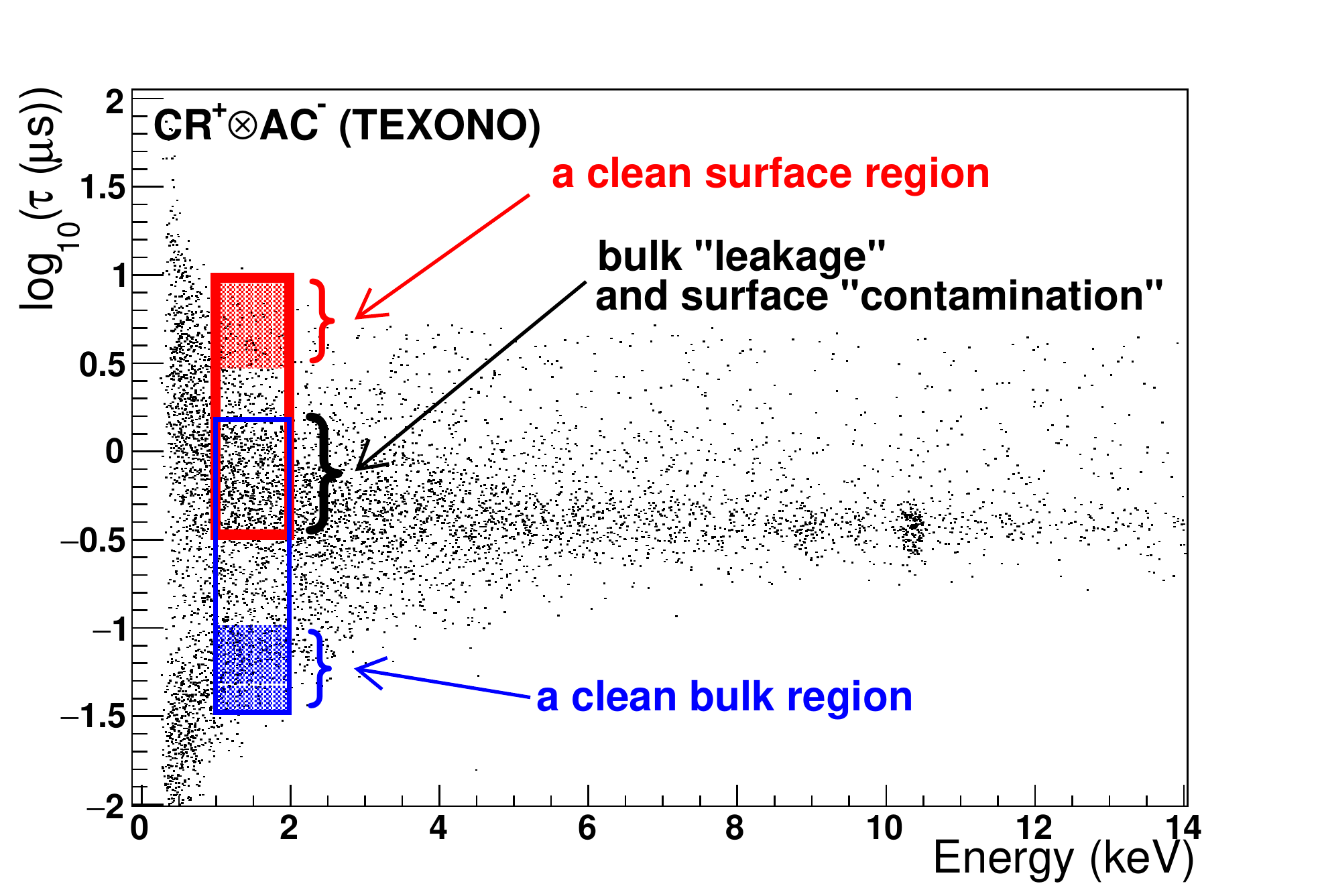}
  \caption{}
\end{subfigure}
\caption{
Rise-time versus energy plot for TEXONO data~\cite{Li:2013a}:
(a) anti-Compton and cosmic-ray vetoed samples
(${\rm{CR}}^{-}\otimes{\rm{AC}}^{-}$),
corresponding to neutrino- and WIMP-induced candidates, and 
(b) cosmic-ray coincident and anti-Compton vetoed samples 
(${\rm{CR}}^{+}\otimes{\rm{AC}}^{-}$), 
corresponding to cosmic-ray induced high-energy neutron
interactions.
The shaded blue and red regions correspond to the
parameter space in B- and S-events, respectively,
where there are no cross-contaminations
among the two samples
such that Eq.~\ref{eq:ratio_scaling_0}  is satisfied.
These regions provide the solution of Eq.~\ref{eq:chi2_fB_fS}.
The overlap of the two boxes
correspond to regions with cross-contaminations. 
}
\label{fig:k62_vv_tv_demo}
\end{figure}

If there exist uncontaminated B- and S- regions
in $\tau$-space, then $\beta_{i}$ and $\xi_{i}$
can be chosen as
\begin{align}
\beta_{i} ( E ) = 
\int_{b_{0}}^{b_{1}}{N_{B i}(E, \tau)d\tau} {\approx}
\int_{b_{0}}^{b_{1}}{N_{i}(E, \tau)d\tau}\nonumber \\
\xi_{i} ( E ) = \int_{s_{0}}^{s_{1}}{N_{S i}( E, \tau)d\tau} {\approx}
\int_{s_{0}}^{s_{1}}{N_{i}(E, \tau)d\tau} ~~ ,
\label{eq:ratio_calculation_0}
\end{align}
as illustrated in Figure~\ref{fig:k62_vv_tv_demo},
with $\tau{\in}[b_{0}, b_{1}]$ as boundaries of blue shadow box, and
$\tau{\in}[s_{0}, s_{1}]$ as boundaries of red shadow box.
This choice of $\beta_{i}$ and $\xi_{i}$ satisfies
Eq.~\ref{eq:common_b_plus_s_0} and \ref{eq:ratio_scaling_0},
and provides the required scaling factors to solve Eq.~\ref{eq:chi2_fB_fS}.
The boundary values ($b_{0}$, $b_{1}$, $s_{0}$ and $s_{1}$)
are $E$-dependent in general.
These can be selected within a range
as long as they enclose the uncontaminated 
B- and S-regions.

At low energy near detector threshold, 
there are cross-contaminations between the B- and S-events.
The algorithm to derive the scaling factors $\beta_{i} (E)$ and $\xi_{i} (E)$
in these regions is described in 
Section~\ref{subsection:evaluation}. 

In the limiting case of only two data samples (indexed as 0 and 1),
the solution for $\beta_{0} (E) f_{B}(E, \tau)$ is:
\begin{equation} \label{eq:two_sources_solution}
\beta_{0} (E) f_{B}(E , \tau) =
\frac{N_{1}(E, \tau) - [ \xi_{1}(E) / \xi_{0}(E) ] N_{0}(E, \tau)}
 { [ \beta_{1}(E) / \beta_{0} (E)] - [ \xi_{1}(E) / \xi_{0}(E) ] } .
\end{equation}
The solution is undetermined 
at $\beta_{1}/\beta_{0}=\xi_{1}/\xi_{0}$.
That is, splitting a data set 
into two each having the same rise-time distribution profile
would not provide a solution.
The solutions for $\beta_{i}(E) f_{B}(E, \tau)$ and 
$\xi_{i}(E) f_{S}(E, \tau)$ exist
only if at least two of the sources satisfy 
$\beta_{i}/\beta_{j}{\neq}\xi_{i}/\xi_{j}$.
When all the sources have same $\beta_{i}/\beta_{j}$ and $\xi_{i}/\xi_{j}$,
the statistic uncertainty will approach infinity (i. e., denominator 
of Eq.~\ref{eq:two_sources_solution} approaches zero).

Discussions on statistical and systematic uncertainties
of this algorithm are discussed in Section~\ref{subsection::uncertainties}
in connection with the analysis on experimental data.

\subsection{Cross-Contamination Regions}\label{subsection:evaluation}

As illustrate in the $E < 2 ~ {\rm keV}$ range 
in Figure~\ref{fig:rt_vs_lowE}, 
there are contamination of S-events 
into $[ b_{0}, b_{1} ]$ and of B-events 
into $[ s_{0}, s_{1} ]$.  
In these domains, 
$\beta_{i} (E)$ and $\xi_{i} (E)$ could be 
derived by a successive approximation algorithm
formulated as: 
\begin{align}
\beta^{(n)}_{i} (E) = \beta^{0}_{i}  (E) - 
  \int_{b_{0}}^{b_{1}} 
   {\xi^{(n-1)}_{i} (E)  ~ f^{(n-1)}_{S}(E, \tau) ~ d\tau} \nonumber \\
\xi^{(n)}_{i} (E) = \xi^{0}_{i} (E) -
  \int_{s_{0}}^{s_{1}}
      {\beta^{(n-1)}_{i} (E)  ~ f^{(n-1)}_{B}(E, \tau) ~ d\tau}, 
\label{eq:ratio_correction_n}
\end{align}
where $\beta^{0}_{i} (E) $ and $\xi^{0}_{i} (E) $ are initial guesses
of scaling factors evaluated from Eq.~\ref{eq:ratio_calculation_0},
and $f^{(n-1)}_{B} (E, \tau)$ and $f^{(n-1)}_{S} (E, \tau)$ 
are results of minimizing Eq.~\ref{eq:chi2_fB_fS} 
in the ${(n-1)}^{th}$-iteration.

At convergence for large $n$,
the real B- and S-event rates for the $i^{th}$-samples are:
\begin{eqnarray}
B_{r i}(E) & = & \int_{all~\tau}N_{B i}(E,\tau)~ d\tau =
    \int_{all~\tau}\beta^{(n)}_{i}(E) ~ f^{(n)}_{B}(E,\tau) ~ d\tau 
~~ {\rm and } ~~
\nonumber \\
S_{r i}(E) & = & \int_{all~\tau}N_{S i}(E,\tau) ~ d\tau =
    \int_{all~\tau}\xi^{(n)}_{i}(E) ~ f^{(n)}_{S}(E,\tau) ~ d\tau ~~ , 
  \label{eq:Br_Sr_calculation} 
\end{eqnarray}
respectively.

In practice,
we adopted a 10-iteration calculation in this analysis.
A systematic cross-check was performed with a 100-iteration
calculation, where the difference is less than 0.01\%.

\section{Data Analysis}\label{section:on_c1a_data}

Published data from the CDEX-1 experiment~\cite{Yue:2014a,Zhao:2016} 
were analyzed using the ratio method, 
and the results were compared with the results from 
the spectral shape method. 
Additional consistency checks were performed with
TEXONO data~\cite{Li:2013a,Li:2014a}. 

In both cases, the same event selections prior to BSD were made,
including rejection of events due to electronic noise,
and in coincidence with the cosmic-ray or anti-Compton detectors.
In particular, 
events with extreme slow rise-time ($\tau>10~{\mu}s$)
were discarded, since the contaminations of
B-events to this region is negligible.
These extremely large $\tau$ events were added 
to the S-samples to give the final $S_{r i} (E)$.

\subsection{Rise-time Uniformity and Calibration Samples}\label{section:uniformity_and_calibration}

\begin{figure}
  \centering
  \includegraphics[width=0.8\linewidth]{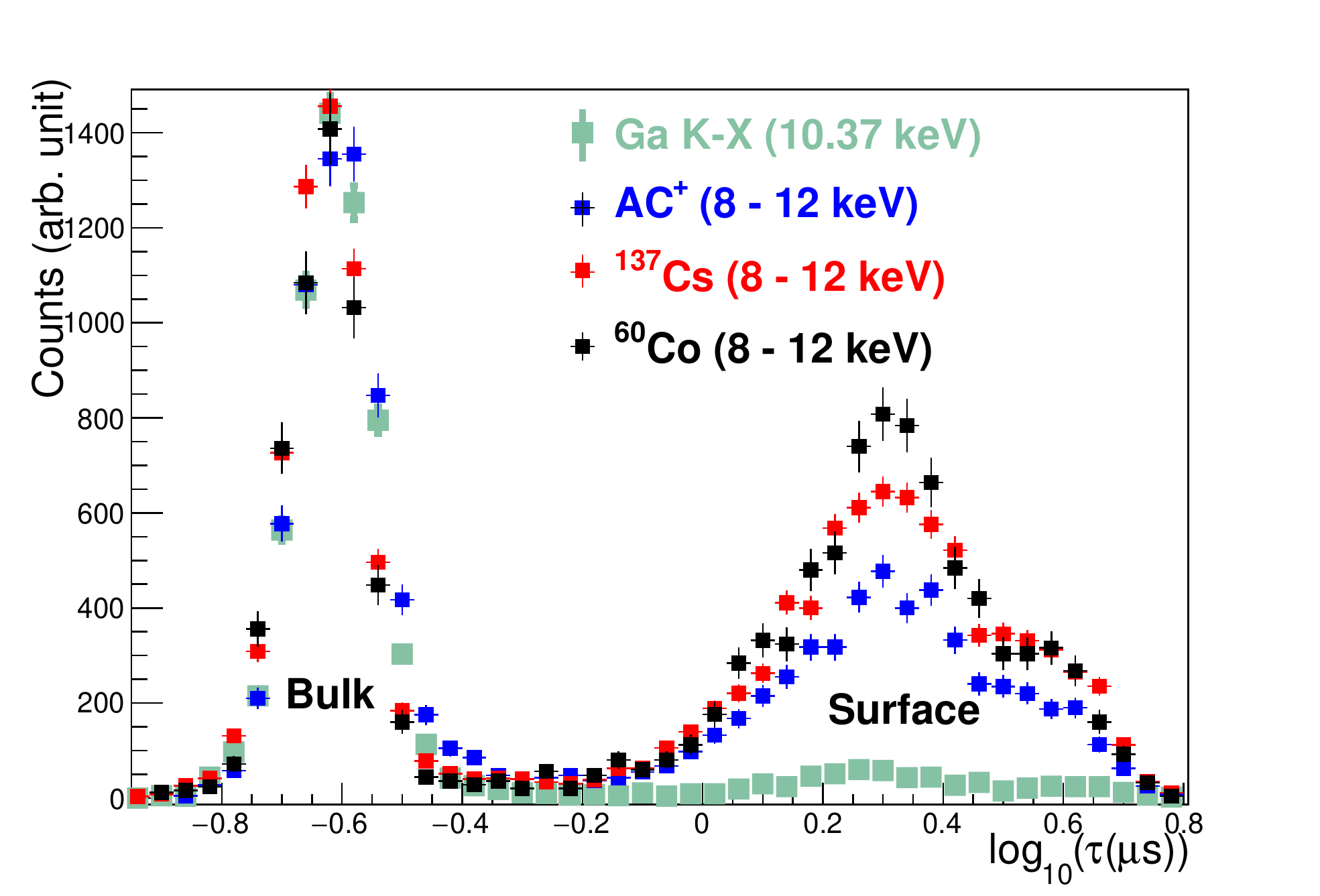}
\caption{Rise-time distributions of Ga K-shell X-rays,
$\rm{AC}^{+}$, $^{137}\rm{Cs}$ and $^{60}\rm{Co}$ events
from CDEX-1 experiment at 8$-$12~keV.
}
\label{fig:normalize_tau_distribution_800_1200}
\end{figure}

\begin{figure}
\centering
\begin{subfigure}{0.5\textwidth}
  \centering
  \includegraphics[width=1.08\linewidth]{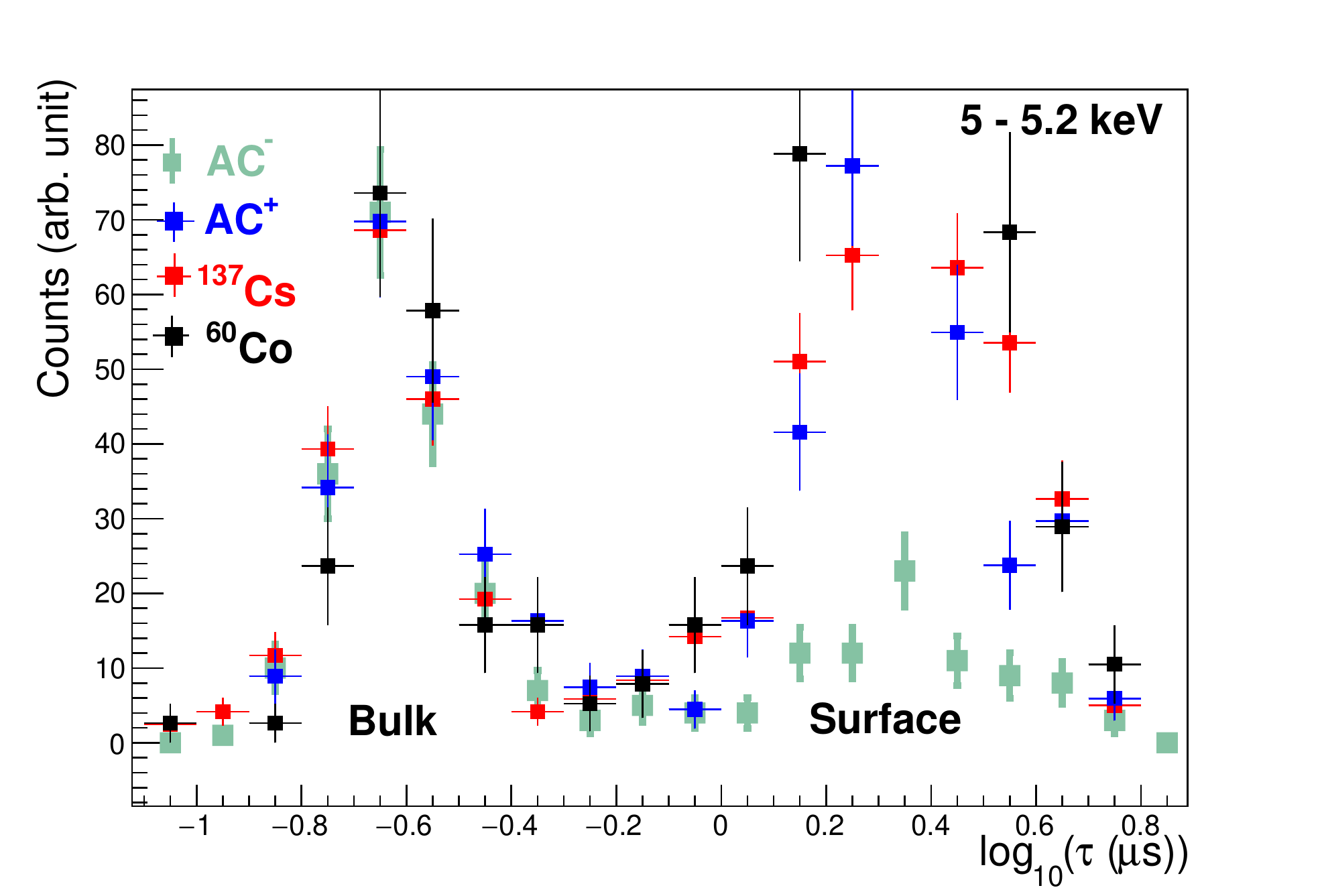}
  \caption{}
\end{subfigure}%
\begin{subfigure}{0.5\textwidth}
  \centering
  \includegraphics[width=1.08\linewidth]{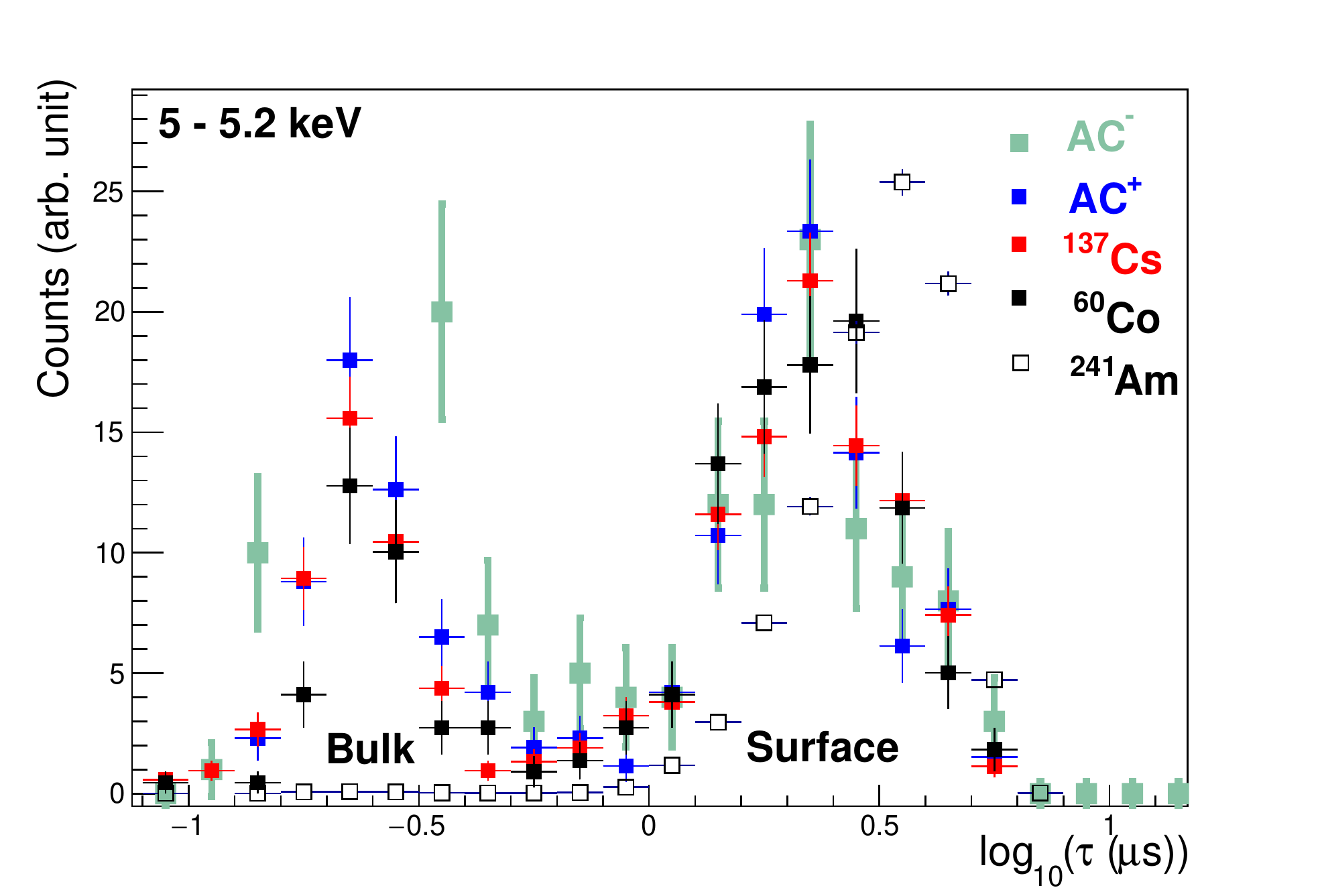}
  \caption{}
\end{subfigure}
\caption{
Rise-time distributions at 5$-$5.2~keV, (a) normalized to bulk counts and
(b) normalized to surface counts show that B/S distributions for
${\rm{AC^{-}}}$, ${\rm{AC^{+}}}$, $^{137}{\rm{Cs}}$ and $^{60}{\rm{Co}}$
are consistent,
with $^{241}{\rm{Am}}$ (which is not used in the analysis) for
comparison.
Note that at 5~keV, bulk and surface events are well
separated.
}
\label{fig:normalize_tau_distribution5052}
\end{figure}

\begin{figure}
\centering
\begin{subfigure}{0.5\textwidth}
  \centering
  \includegraphics[width=1.08\linewidth]{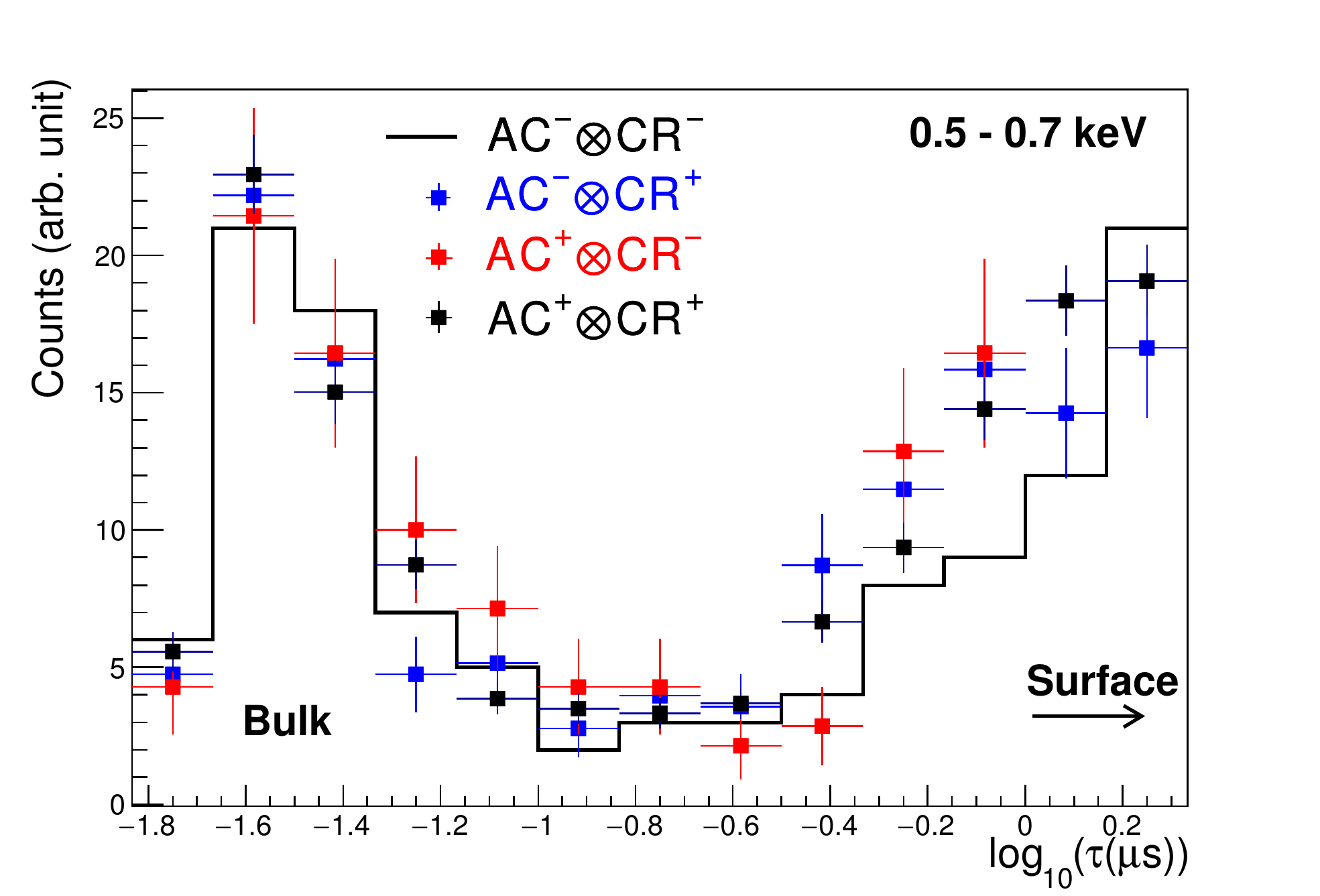}
  \caption{}
\end{subfigure}%
\begin{subfigure}{0.5\textwidth}
  \centering
  \includegraphics[width=1.08\linewidth]{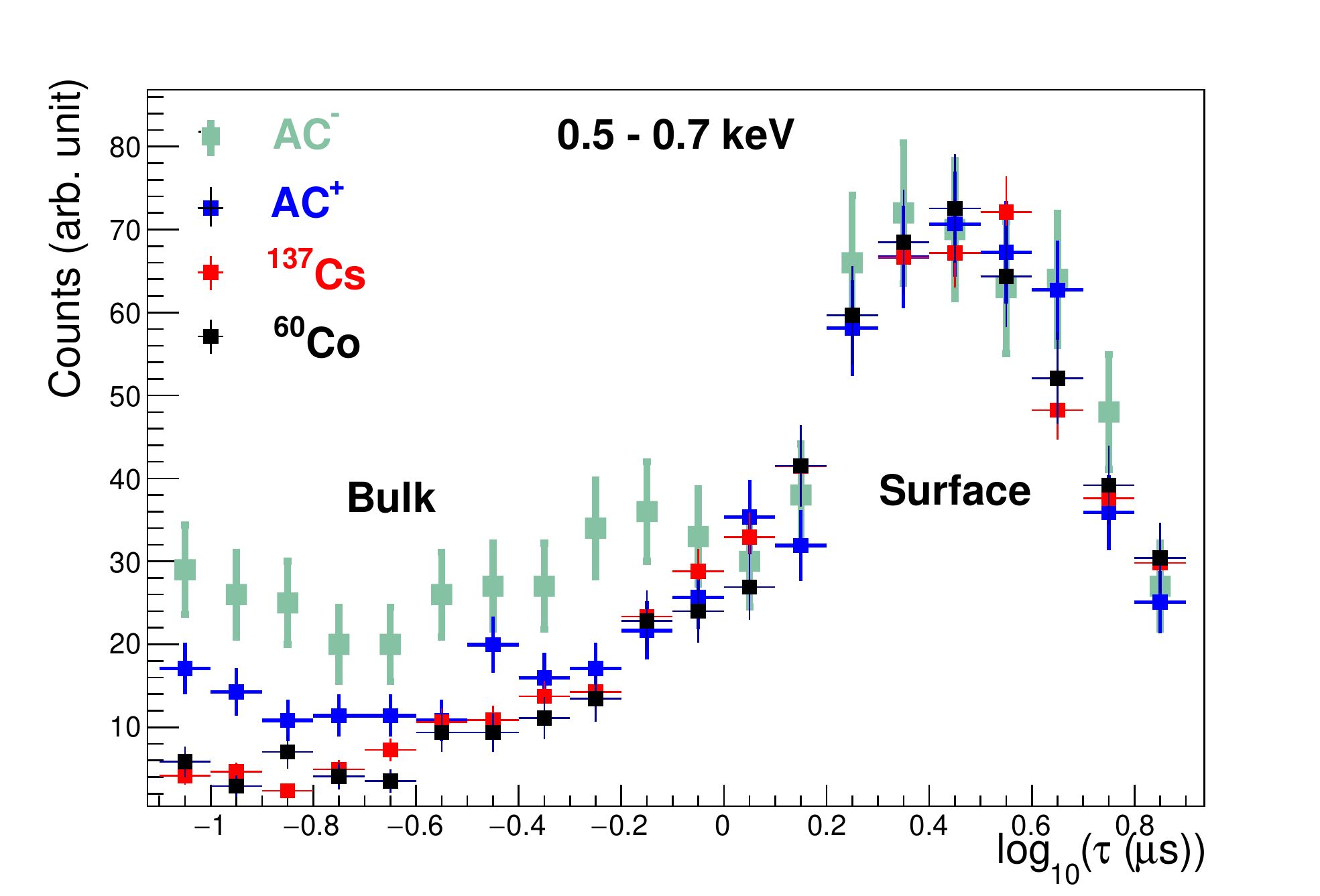}
  \caption{}
\end{subfigure}
\caption{(a) Rise-time distributions of ${\rm{AC^{\pm}{\otimes}CR^{\pm}}}$ from TEXONO~\cite{Li:2013a,Li:2014a}
at 0.5$-$0.7~keV,
showing that ${\rm{AC^{-}{\otimes}CR^{+}}}$ events are consistent with the
others which are surface rich due to external $\gamma$-rays.
(b) Rise-time distributions at 0.5$-$0.7~keV show that surface distributions for all sources are consistent.
}
\label{fig:normalize_tau_distribution0507}
\end{figure}

\begin{figure}
  \centering
  \includegraphics[width=0.8\linewidth]{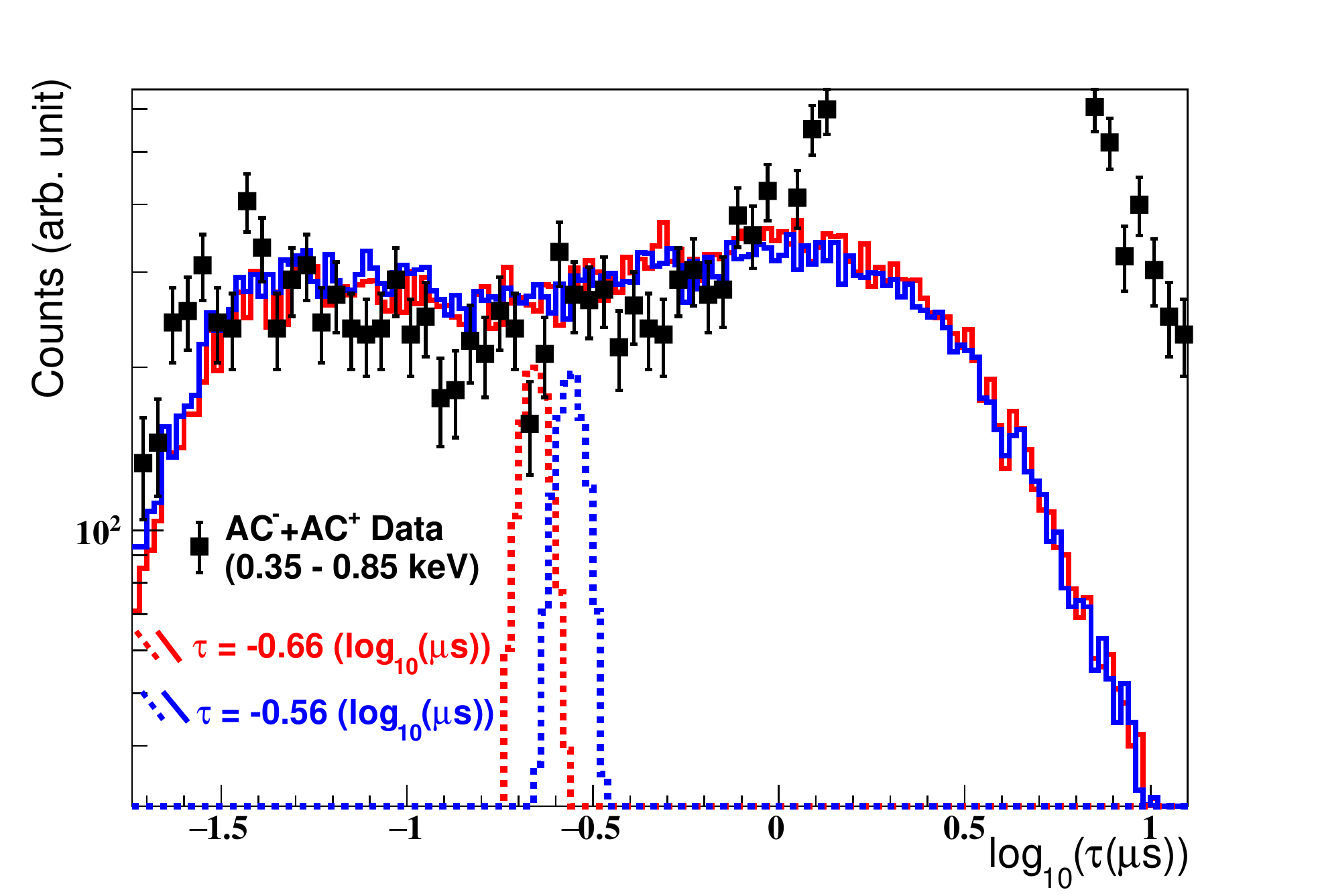}
\caption{
Comparison of rise-time distributions of $\rm{AC}^{-}+{AC}^{+}$ data in
CDEX-1 (0.35$-$0.85~keV)
with those from simulated pulses at two different rise-times shifted by
0.1~(${\rm{log_{10}({\mu}s)}}$).
The dotted and solid lines correspond to those derived from amplitude
of 10.37~keV and 0.6~keV, respectively, coupled with random pedestal noise.
}
\label{fig:noise_sim_0507}
\end{figure}

%% Table
\begin{table}[h!]
\centering
\resizebox{\textwidth}{!}{
 \begin{tabular}{|l c c c c |c c|}
 \hline
 & 5$-$5.2~keV & 5$-$5.2~keV & 0.5$-$0.7~keV
     & 0.5$-$0.7~keV & 10.37~keV & 0.5$-$0.7~keV \\
 & surface & bulk & surface
     & bulk & bulk & bulk \\
$\tau$ range (${\rm{log_{10}({\mu}s)}}$) &  0.1$-$0.45
     & {\rm{-}}0.75$-${\rm{-}}0.5 & 0.2$-$0.8
     & {\rm{-}}1.8$-${\rm{-}}1.2 & simulation & simulation \\
 & & & & (TEXONO) & & \\
 \hline
Mean of $\rm{AC}^{-}$ & 0.36 & {\rm{-}}0.63 & 0.46 & {\rm{-}}1.5 & {\rm{-}}0.46 & {\rm{-}}0.19 \\
$\sigma$ of $\rm{AC}^{-}$ & 0.028 & 0.013 & 0.02 & 0.16 & 0.049 & 0.54 \\
Deviations of mean & 0.03 & 0.02 & 0.02 & 0.03 & 0.1$^\dagger$ & 0.008 \\
 \hline
\multicolumn{7}{l}{$^\dagger$ Input shift for simulation pulses.} \\
 \end{tabular}
}
\caption{
The mean and $\sigma$ of rise-time distributions for bulk and surface
events at high (5.0$-$5.2~keV) and low (0.5$-$0.7~keV) energy, and the maximal
deviations of different samples from the mean as in
Figure~\ref{fig:normalize_tau_distribution5052}\&\ref{fig:normalize_tau_distribution0507}.
An additional estimate is made for the low energy bulk samples (rightmost column) from
the differences in $\tau$ of the simulated pulses of
Figure~\ref{fig:noise_sim_0507}. The maximal deviations are adopted as input
to one of the terms in the evaluation of systematic uncertainties in
Table~\ref{table::error}.
}
\label{table::BS_means}
\end{table}

The validity of this analysis requires calibration source data with consistent
rise-time distributions. These conditions are satisfied automatically
for the B-samples at all energies, as discussed in Section~\ref{section::risetime} and 
shown in Figures~\ref{fig:normalize_tau_distribution_800_1200}, 
\ref{fig:normalize_tau_distribution5052}a 
and \ref{fig:normalize_tau_distribution0507}a.

As depicted in Figure~\ref{fig:normalize_tau_distribution5052}b for the S-samples at keV energy, 
we selected those calibration sources which give consistent rise-times as the physics samples, 
all of which originate from high energy gamma-interactions. 
On the contrary, low energy gamma's from $^{241}{\rm{Am}}$ which have severe attenuation 
at the surface layers cannot be used. 
The optimal selection of the calibration data is different for different experiments.
For {\it this} analysis, samples from $\rm{AC}^{-}$, $\rm{AC}^{+}$,
$^{137}{\rm{Cs}}$ and  $^{60}{\rm{Co}}$
are selected for calibration of the CDEX-1 data~\cite{Yue:2014a,Zhao:2016}
discussed in Section~\ref{section:cdex1_energy_spectra}, and from 
$\rm{CR}^{-}\otimes\rm{AC}^{-}$,
$\rm{CR}^{+}\otimes\rm{AC}^{-}$, $\rm{CR}^{-}\otimes\rm{AC}^{+}$ and
$\rm{CR}^{+}\otimes\rm{AC}^{+}$ of the 
TEXONO data~\cite{Li:2013a} discussed in Section~\ref{section:texono_data}.

At low energies (below 1 keV for the data discussed in this article),
resolution effects smear out the intrinsic rise-time differences for the S-events,
such that the measured rise-time distributions are the same for all sources,
as shown in Figures~\ref{fig:normalize_tau_distribution0507}b and 
\ref{fig:noise_sim_0507}.

The uniformity of rise-time distributions and their independence to 
locations and nature of interactions are demonstrated for the selected calibration samples. 
The $^{137}{\rm{Cs}}$ and $\rm{AC}^{+}$ events
are electron-recoils induced by $\gamma$-rays external to the detector and therefore
have higher probability of located close to the surface. The Ga K-shell X-rays
(10.37 keV) are also electron-recoils but due to cosmogenic activation
inside the detector and are therefore uniformly distributed within the entire
fiducial volume. The $\rm{CR}^{+}\otimes\rm{AC}^{-}$ samples select cosmic-ray induced high energy
neutrons giving rise nuclear recoil events at the detector. Both the
energy distribution (exponential rise towards low energy) and bulk-surface
events ratio (uniformly distributed with detector) show these selected samples
are neutron-rich~\cite{Li:2014a}. By comparing with neutron flux measurement
with a hybrid liquid scintillator detector~\cite{Singh:2017} placed at the same location as
the Ge-target, the fraction of nuclear recoils is about 99\%~\cite{Sonay:2017}. 
The measured
bulk-event rise-time distributions for these samples 
($^{137}{\rm{Cs}}$, $\rm{AC}^{+}$, Ga X-rays, $\rm{CR}^{+}\otimes\rm{AC}^{-}$) 
are all consistent with each other.

Gaussian fits are performed to derive the mean and root-mean-square
(RMS) of the rise-time distributions. The $\rm{AC}^{-}$ samples are candidate events
uncorrelated with other detector components and therefore the subjects of
physics analysis. The deviations of various sources relative to the $\rm{AC}^{-}$ events
are summarized in Table~\ref{table::BS_means}. The maximal shift of the mean is $\sim$~1~RMS. This
deviation matches the expectations due to measurement and statistical uncertainties,
and has been taken in account in the consideration of systematic
uncertainties to be discussed in Section~\ref{subsection::uncertainties}.

Analysis on simulated pulses is performed to provide additional support to
the rise-time independence. Rise-time of 10.37~keV
Ga K-shell X-rays events were measured. Two event samples with $\tau$ at $\pm$~1~RMS of the 
mean were extracted and added together to obtain their respective ``averaged" pulse shape. 
These smoothed reference pulses were
added to a large sample of random pedestal noise profiles. The measured
rise-time of these simulated events are depicted in Figure~\ref{fig:noise_sim_0507}. It shows two
Gaussians with $\Delta\tau$ = 0.1~(${\rm{log_{10}({\mu}s)}}$). 
The same analysis was repeated with
the reference pulses scaled to 0.6~keV instead. The measured rise-time distributions,
also displayed in Figure~\ref{fig:noise_sim_0507}, show broad profiles identical in both samples. 
As listed in Table~\ref{table::BS_means}, the corresponding shift of the means is less
than the RMS, demonstrating that an artificial shift of intrinsic rise-time
would produce no measureable effects at low energy, which is the crucial
region of interest in BSD analysis. This further justifies the validity of the calibration 
samples selection.

\subsection{Best-fit of Rise-time Distributions}\label{section:cdex_1a_results}

\begin{figure}[!htbp]
\centering\includegraphics[width=1.0\linewidth]{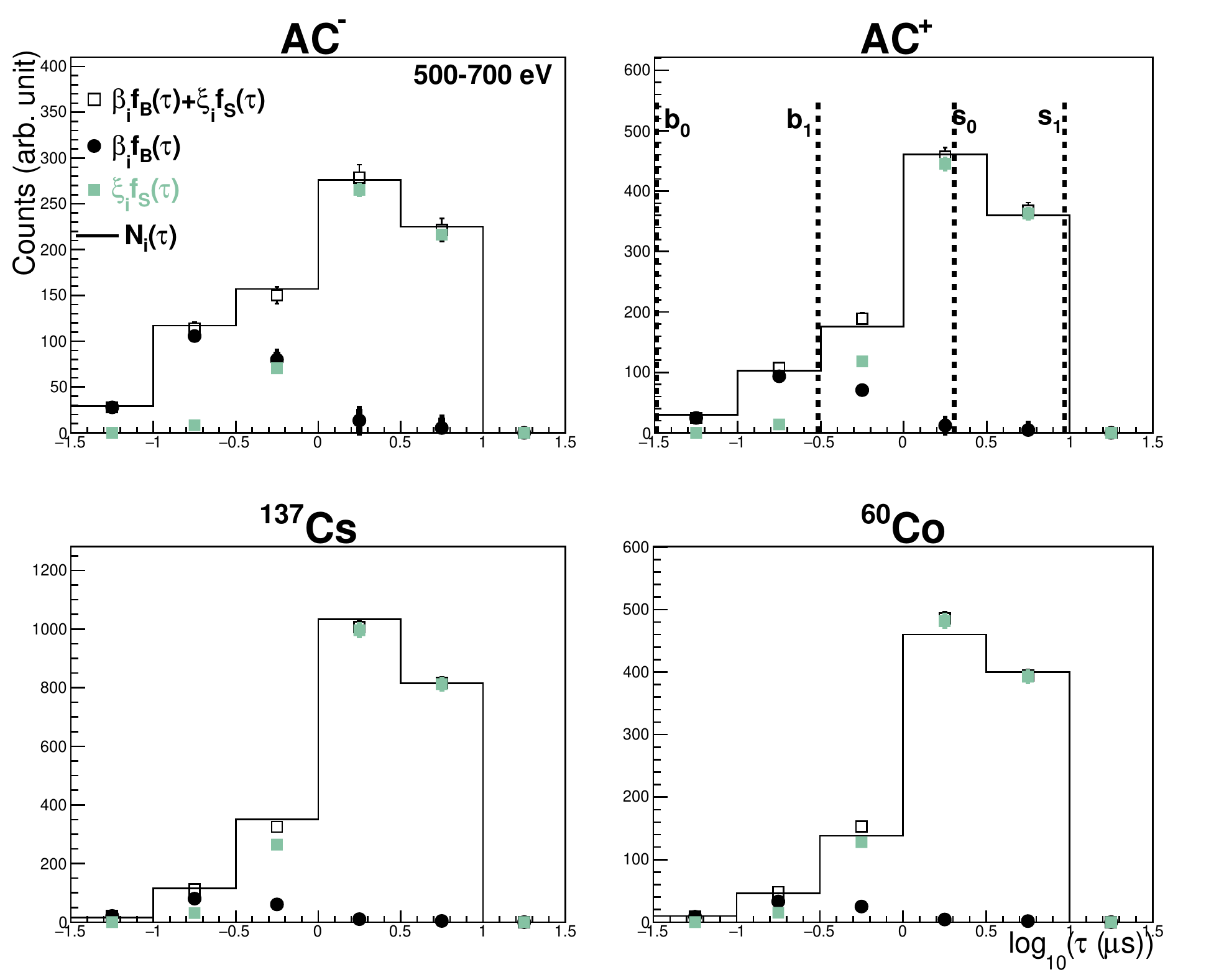}
\caption{ 
Best-fit results of Eq.~\ref{eq:chi2_fB_fS} on
CDEX-1 data,
for $f_{B}(E, \tau)$ and $f_{S}(E, \tau)$ at
$E = 500$-$700~{\rm eV}$. The histogram  $N_{i}(E, \tau)$ corresponds
to the raw data, with
$i~=~\{\rm{AC}^{-}, \rm{AC}^{+}, ^{137}\rm{Cs}, ^{60}\rm{Co} \}$. 
The total $\chi^{2}$/dof for the combined data is 10.5/10.
}
\label{fig:best_fit_c1a_500_700eV}
\end{figure}

Two {\it in situ} event samples are available in the CDEX-1 data:
$\rm{AC}^{-}$ which are the WIMP candidate events and 
$\rm{AC}^{+}$ which are background due to ambient radioactivity.
In addition, calibration data from
$^{137}\rm{Cs}$ and $^{60}\rm{Co}$ sources were also taken.
As discussed in Section~\ref{section::risetime}, these data samples have 
similar $f_B (E, \tau)$ and $f_S (E, \tau)$ distributions at large $E$ where
the BSD is distinct, but they also complement
each other through having
different B- and S-events ratios. 

Contrary to the spectral shape method,
the ratio method does not require  
assumption or simulation input on spectral shape.
Accordingly, all four samples can contribute to BSD.
Although two of these are in principle sufficient to provide
solutions to Eq.~\ref{eq:chi2_fB_fS}, 
the information from all samples would
provide redundancy and reduce uncertainties
especially at low energy ($<$1~keV).

Best fit results for 
$f_{B} (E, \tau)$ and $f_{S} (E, \tau)$ at 
$E = 500$-$700~{\rm eV}$
are depicted in Figure~\ref{fig:best_fit_c1a_500_700eV}.

\subsection{Uncertainties and Goodness-of-Fit}
\label{subsection::uncertainties}

The sources and size 
of statistical and systematic uncertainties
on a low and a high energy bins 
of the CDEX-1 $B_{r i}$ samples at the
$i=\rm{AC}^{-}$ channel are summarized
in Table~\ref{table::error}.
Standard error propagation techniques are used to evaluate the
combined uncertainties
of $B_{r i}$ and $S_{r i}$ for every $E$-bin.

\begin{figure}[!htbp]
\centering\includegraphics[width=0.8\linewidth]{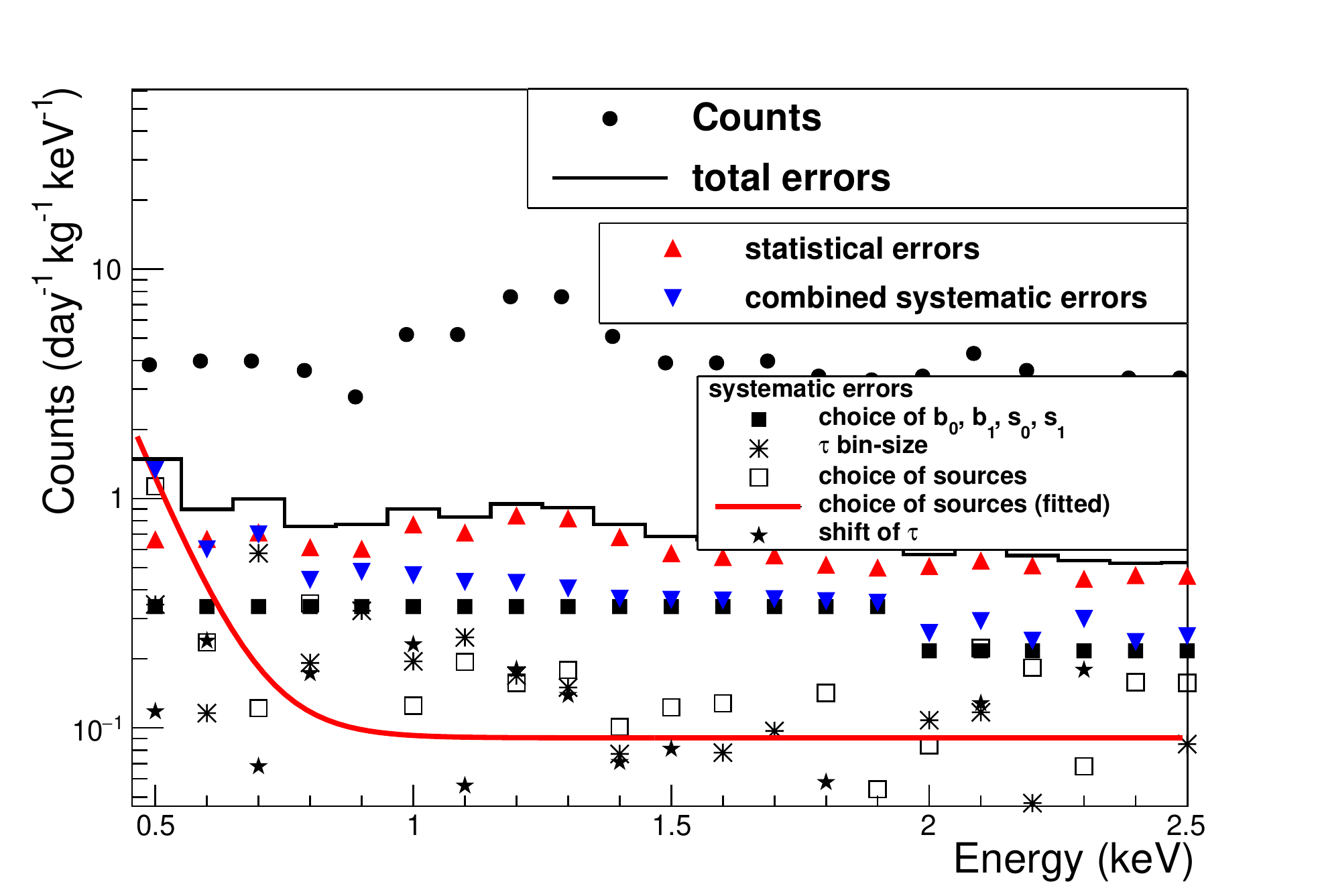}
\caption{
Distribution of count rates and their uncertainties
of $B_{r i} ( E )$ sample of CDEX-1 data
for the  $i = {\rm AC^-}$ channel. 
The various contributions to both statistical and systematic errors 
are also shown.
The threshold bin is with $E = 450$$-$$550~{\rm eV}$.
At low energy, the largest contribution comes from 
the choice of sources and
the choice of $b_{0}$, $b_{1}$, $s_{0}$ and $s_{1}$, 
whereas errors are dominated by statistical errors
at high energy.
}
\label{fig:c1a_errors_budget}
\end{figure}

There are three factors contributing to the statistical errors:
\begin{enumerate}
\item Errors at 1-$\sigma$ level on $f_{B}$ and $f_{S}$
are calculated from $\chi_{min}^{2}+1$
      of Eq.~\ref{eq:chi2_fB_fS} for each $(E, \tau)$-bin;
\item Errors of $\beta^{0}_{i}$ and $\xi^{0}_{i}$ for every $E$-bin;
\item Errors of correction terms of $\beta^{(n-1)}_{i}$ and $\xi^{(n-1)}_{i}$
      in Eq.~\ref{eq:ratio_correction_n}.
\end{enumerate}

%% Table
\begin{table}[h!]
\centering
 \begin{tabular}{|l c c|}
 \hline
 Energy & 0.45$-$0.55~keV & 0.55$-$0.65~keV \\
Counts and errors & 3.83$\pm$0.66[stat]
                  & 3.99$\pm$0.67[stat] \\
(${\rm{day}}^{-1}$${\rm{kg}}^{-1}$${\rm{keV}}^{-1}$) & $\pm$1.23[sys] & $\pm$0.60[sys] \\
 \hline
Systematic uncertainties: & & \\
(1) Choice of $b_{0}$, $b_{1}$, $s_{0}$ and $s_{1}$: & 0.34 & 0.34  \\
(2) $\tau$ bin-size: & 0.35 & 0.12 \\
(3) Choice of sources: & 1.12 & 0.42 \\
(4) Shift of $\tau$ by 0.02~(${\rm{log_{10}({\mu}s)}}$): & 0.12 & 0.24 \\
%% ADD v22
(5) Contribution of low energy $\gamma$: & $<$0.0028 & $<$0.0028 \\
(6) Non-zero counts at clean-bulk/surface: & $<$0.038 & $<$0.038 \\
(7) Iterations of $\beta_{i} (E)$, $\xi_{i} (E)$: & $<{\rm{10^{-4}}}$ & $<{\rm{10^{-4}}}$ \\
 \hline
 \end{tabular}
\caption{The various contributions to 
the systematic uncertainties of $B_{r}$ of AC$^-$ at two different energy bins.
}
\label{table::error}
\end{table}

\begin{figure}[!htbp]
\centering\includegraphics[width=1.0\linewidth]{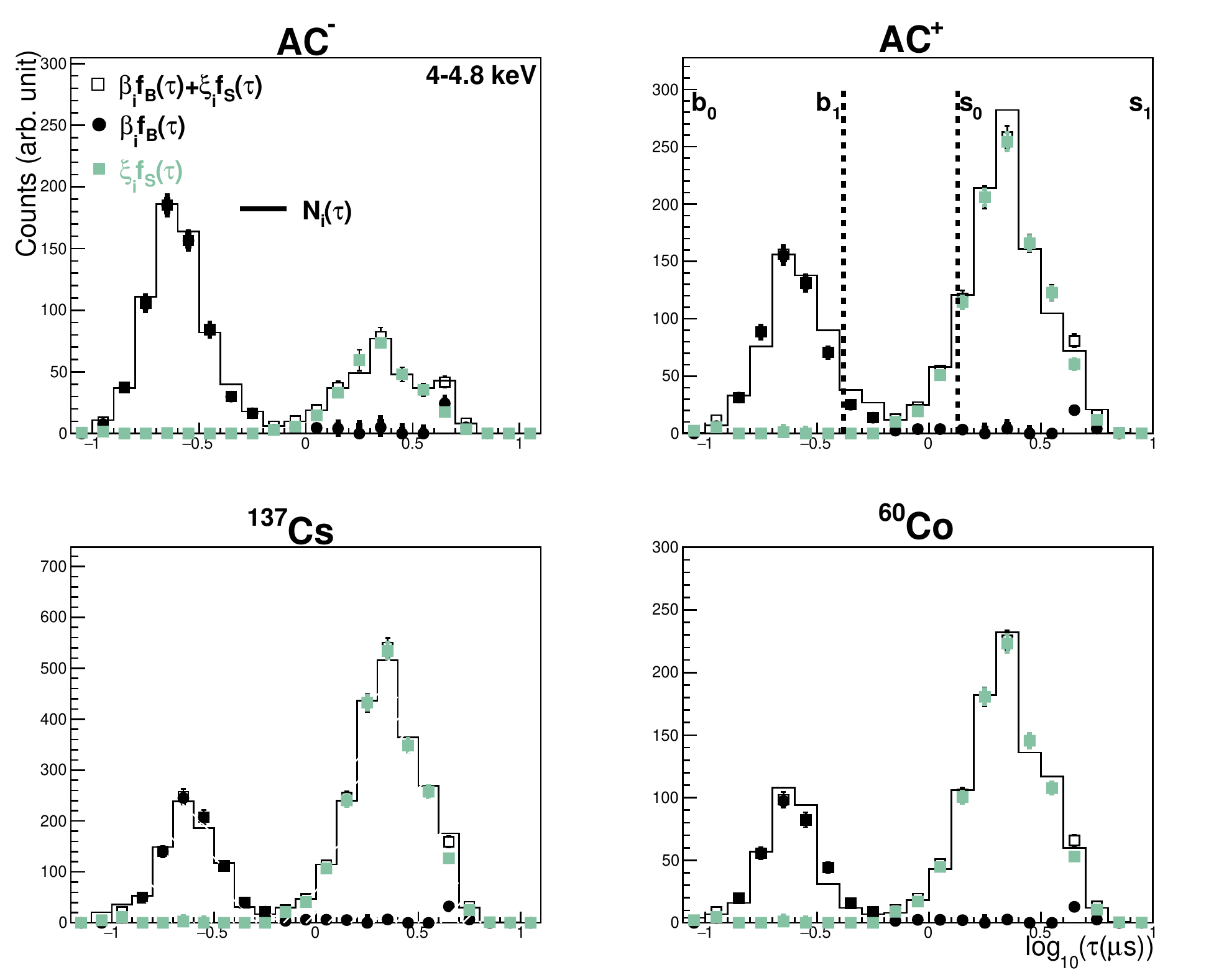}
\caption{
Best-fit results of Eq.~\ref{eq:chi2_fB_fS} on
CDEX-1 data,
for $f_{B}(E, \tau)$ and $f_{S}(E, \tau)$ at
$E = 4$-$4.8~{\rm keV}$. The histogram  $N_{i}(E, \tau)$ corresponds
to the raw data, with
$i~=~\{\rm{AC}^{-}, \rm{AC}^{+}, ^{137}\rm{Cs}, ^{60}\rm{Co} \}$.
A finer binning is used to demonstrate that $f_{B}$ is close to zero at
clean-surface region (and vice versa). Small amount of non-zero
$f_{B}$ counts at clean-surface region is caused by statistical non-uniformity.
Nevertheless, those non-zero counts could serve as
measurement of non-uniformity.
}
\label{fig:best_fit_c1a_400_480eV}
\end{figure}

%%modify v24
The systematic uncertainties of the $E = 450$$-$$550~{\rm eV}$ bin are displayed 
in Figure~\ref{fig:c1a_errors_budget}. Their derivations are discussed as follows, 
in which items 3-6 are related
to possible non-uniformities of the rise-time pulse shape:
\begin{enumerate} [(1)]
\item \textbf{Choice of $b_{0}$, $b_{1}$, $s_{0}$, $s_{1}$.} \\
The ranges of [$b_{0}$, $b_{1}$] and [$s_{0}$, $s_{1}$]
according to Figure~\ref{fig:rt_vs_lowE}
are reduced by 25\%, and the maximum deviations in the results within a
2~keV bin are taken as systematic errors.
Reduced ranges imply larger fluctuations in the count rates.

\item \textbf{Choice of $\tau$ bin-size.} \\
Systematic effects are taken as deviations of results due to variations of 
bin-size, from half to twice the nominal one.

\item \textbf{Choice of different combinations of calibration data.} \\
This is the largest contribution of systematic uncertainties
at energy near threshold.
Identical analysis were performed without
${\rm{AC^{+}}}$, $^{137}\rm{Cs}$ or $^{60}\rm{Co}$,
one at a time. The systematic uncertainties are assigned from the 
best-fit function
(the red curve of Figure~\ref{fig:c1a_errors_budget}) of the maximum deviations.
The results remain mostly unchanged at high energy where bulk and surface events
are well separated. However, at low energy, where B/S mixture is severe, 
the uncertainty of the B/S separation depends on the number of calibration sources. 
Removing sources increase uncertainties with strong energy dependence, as
shown in Figure~\ref{fig:c1a_errors_budget}.

\item \textbf{Deliberately shifting the mean rise-time of calibration events.} \\
The rise-time of ${\rm{AC^{-}}}$ is shifted by the amount
allowed in Table~\ref{table::BS_means}, and the deviations of results are taken 
as systematic errors.

\item \textbf{Extra low energy $\gamma$'s component at surface region.} \\
Surface rise-time distributions of high energy $\gamma$'s, e. g.,
${\rm{AC^{+}}}$, $^{137}{\rm{Cs}}$ and $^{60}{\rm{Co}}$
resemble that of ${\rm{AC^{-}}}$. Sources that do not resemble ${\rm{AC^{-}}}$
can be represented by low energy $\gamma$'s
from $^{241}{\rm{Am}}$ whose surface rise-time distributions at 5~keV is shown in
Figure~\ref{fig:normalize_tau_distribution5052}b.

Upper bounds of their contributions to systematic uncertainties
could be calculated by a simplified two components linear-fit
to the surface region of ${\rm{AC^{-}}}$  at the high energy region:
\begin{equation}
N_{{\rm{AC^{-}}}}(\tau) =
{\alpha} N_{ {\rm{AC^{+}}}+^{137}{\rm{Cs}}+^{60}{\rm{Co}} }(\tau)
+ (1-\alpha) N_{^{241}{\rm{Am}}}(\tau).
\label{eq::am_limit_fit}
\end{equation}
The best-fit results show that
$(1-\alpha)$$=$7.1$\pm$8.4\% (68\% C. L.),
which corresponds to deviation of $B_{r {\rm{AC^{-}}}}$ by $<$~0.1~counts
(${\rm{day}}^{-1}$${\rm{kg}}^{-1}$${\rm{keV}}^{-1}$)
or $<$~0.2\% increasing in total errors at 450$-$550~eV.

\item \textbf{Finite $f_{B}$ ($f_{S}$) counts at clean-surface (-bulk) region.} \\
In ideal cases, the $f_{B}(\tau)$ (or $f_{S}(\tau)$) solution
should perfectly match with the rise-time distributions of the sources,
and $f_{S}(\tau)$ (or $f_{B}(\tau)$) should be exactly zero in the clean-bulk 
(or clean-surface) region at 3$-$12~keV.
Therefore, the finite $f_{B}$ ($f_{S}$) counts measurements can be served to quantify 
non-uniformities among chosen sources.

Figure~\ref{fig:best_fit_c1a_400_480eV} depicts the best-fit results
at 4$-$4.8~keV. Finite $f_{S}(\tau)$ counts at clean-bulk region
(and $f_{B}(\tau)$ counts at clean-surface region) contributes to
$<$~0.24~counts (${\rm{day}}^{-1}$${\rm{kg}}^{-1}$${\rm{keV}}^{-1}$) 
of $B_{r {\rm{AC^{-}}}}$. These originates from statistical fluctuations and 
possible pulse-shape non-uniformity.

At low energy, the finiteness originates from B/S contaminations and statistical 
fluctuations, as well as pulse-shape non-uniformity. That provides a measurement 
of upper bounds on the effects due to non-uniformity.
At 500$-$700~eV, contribution of finite $f_{B}$ ($f_{S}$) 
counts is $<$~0.39~${\rm{day}}^{-1}$${\rm{kg}}^{-1}$${\rm{keV}}^{-1}$ of $B_{r {\rm{AC^{-}}}}$,
equivalent to $<$~2.7\% increasing in total errors of $B_{r {\rm{AC^{-}}}}$.

\item Systematic uncertainties from the iterations of $\beta_{i} (E)$, $\xi_{i} (E)$
corrections are negligible.

\end{enumerate}

%%ADD v24

In addition, there are no indications from the literature and from measurements 
that there may be intrinsic pulse shape differences between high and 
low energy B-events. 
The data shows that even large differences 
in the Surface pulse shapes at high recoil energies between high and 
low-energy gamma sources are washed out at low recoil energy.
It is therefore justified that residual differences in the B-event 
pulse shapes, if they exist, would be negligible at low energy.

\begin{figure}[!htbp]
\centering\includegraphics[width=0.8\linewidth]{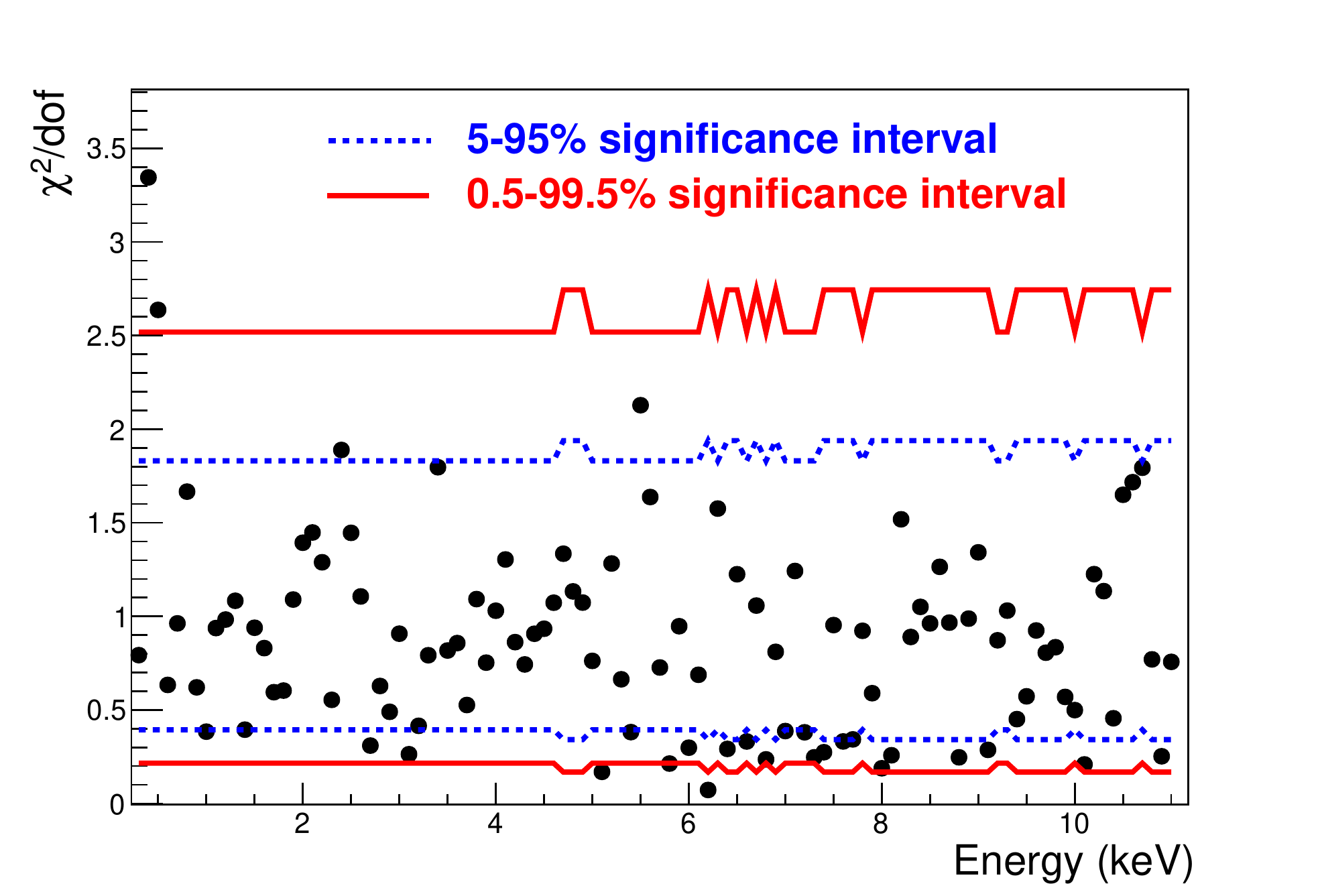}
\caption{
The goodness-of-fit ($\chi^{2}$/dof) derived through minimization of 
Eq.~\ref{eq:chi2_fB_fS}
on CDEX-1 data in Figures~\ref{fig:best_fit_c1a_500_700eV}a, b, c, d.
%% MODIFY v24
The threshold bin is with $E = 450$$-$$550~{\rm eV}$ 
(350$-$450~${\rm eV}$ is also shown) and 
every energy bin is independent.
The related significant intervals are also displayed.
}
\label{fig:reduced_chi2}
\end{figure}

Combining both statistical and systematic uncertainties
for every energy bin, the goodness of fit to
Eq.~\ref{eq:chi2_fB_fS} can be assessed 
via the $\chi^{2}$/dof values.
The results are displayed in Figure~\ref{fig:reduced_chi2}. 
Degree of freedom at each energy bin
is the total number of non-zero 
$\tau$-bins of all four sources subtracting off total
non-zero $\tau$-bins of $f_{B}$ and $f_{S}$. 
The respective significant intervals
are superimposed on Figure~\ref{fig:reduced_chi2},
%%ADD v24
indicating valid fit results and justifying that the four data samples 
share similar rise-time profiles in $f_{B}$ and $f_{S}$
above $\sim$550~eV for this data set~\cite{Zhao:2016}.

%%XXX : still too aggressive, I think.

\subsection{Energy Spectra}
\label{section:cdex1_energy_spectra}

\begin{figure}[!htbp]
\centering\includegraphics[width=0.8\linewidth]{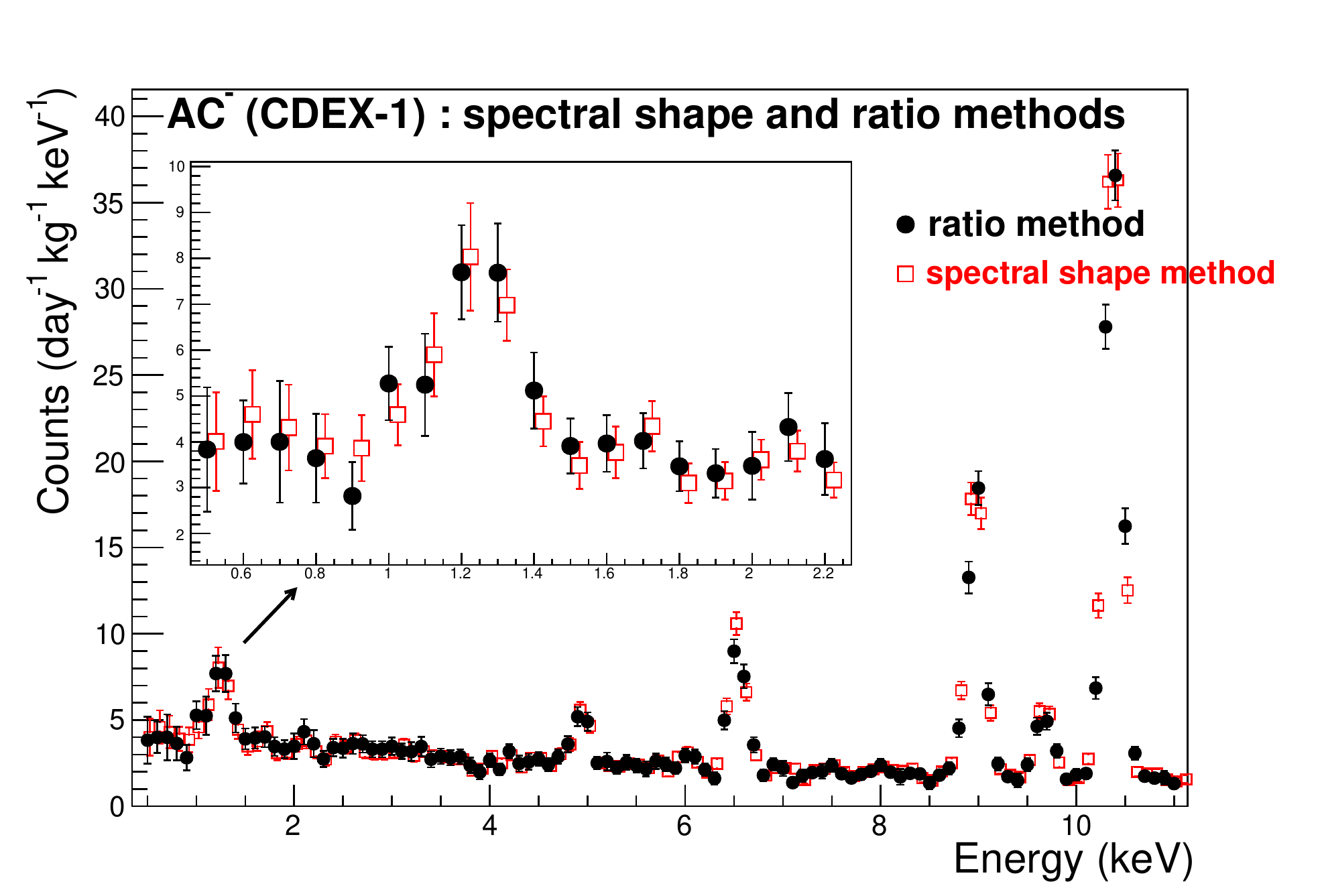}
\caption{
The CDEX-1 $B_{r i} (E)$ spectra for AC$^-$ sample.
Analysis is performed 
and results are compared with both spectral shape and ratio methods.
The thresholds are 450~eV and 475~eV for the ratio 
and spectral shape methods, respectively. 
Systematic errors are included.
}
\label{fig:c1a_old_new_ACVBr}
\end{figure}

\begin{figure}[!htbp]
\centering\includegraphics[width=0.85\linewidth]{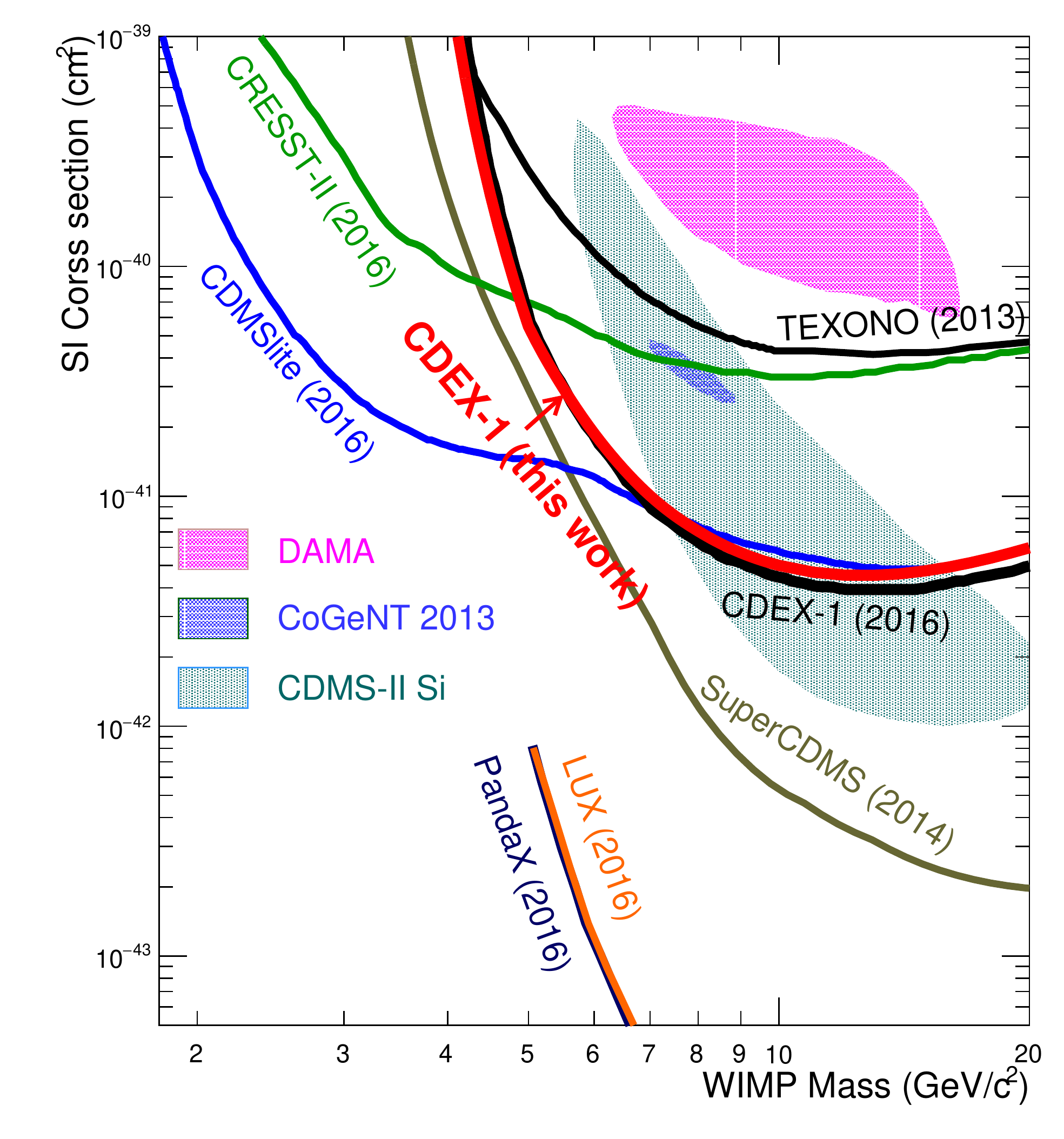}
\caption{
The WIMP-nucleon spin-independent cross-section exclusion plot,
comparing results of the spectral shape method (CDEX-1 (2016)) 
and ratio method (CDEX-1 (this work)) from the CDEX-1 data.
Also shown are 90\% confidence upper limits from 
other benchmark results such as 
CDEX-1 (black)~\cite{Yue:2014a},
CDMSlite (blue)~\cite{Agnese:2016},
CRESST-II (dark green)~\cite{Angloher:2016},
SuperCDMS (olive)~\cite{Agnese:2014},
PandaX (dark blue)~\cite{Tan:2016},
and LUX (orange)~\cite{Akerib:2017}, 
as well as allowed regions  from
DAMA (pink)~\cite{Belli:2011,Bernabei:2010},
CoGeNT (purple)~\cite{Aalseth:2013} and
CDMS II Si(teal)~\cite{Agnese:2013}.
}
\label{fig:c1a_old_new_SI_DM}
\end{figure}

The CDEX-1 energy spectra $B_{r i} (E)$  
at $i=\rm{AC}^{-}$ channel derived with both the
spectral shape and ratio methods are
depicted in Figure~\ref{fig:c1a_old_new_ACVBr},
indicating consistency among them.
The figure also shows that 
all the internal X-ray peaks are correctly reconstructed. 
This is a non-trivial demonstration of validity of the ratio method, 
since every energy bin is processed independently
of the others.

A comparison of the
spin-independent WIMP-nucleon cross-section
exclusion plot 
for both methods is shown in
Figure~\ref{fig:c1a_old_new_SI_DM},
also indicating consistent results.
The slight improvement with the ratio method
at low mass ($<$6~GeV) originates from lower
analyzable threshold (450~eV). 
The slight decrease in sensitivities at high mass is due to increased 
systematic uncertainties when the normalization assumption of the 
previous spectral method is no longer made. 

\begin{figure}
\centering
\begin{subfigure}{0.7\textwidth}
  \centering
  \includegraphics[width=1.08\linewidth]{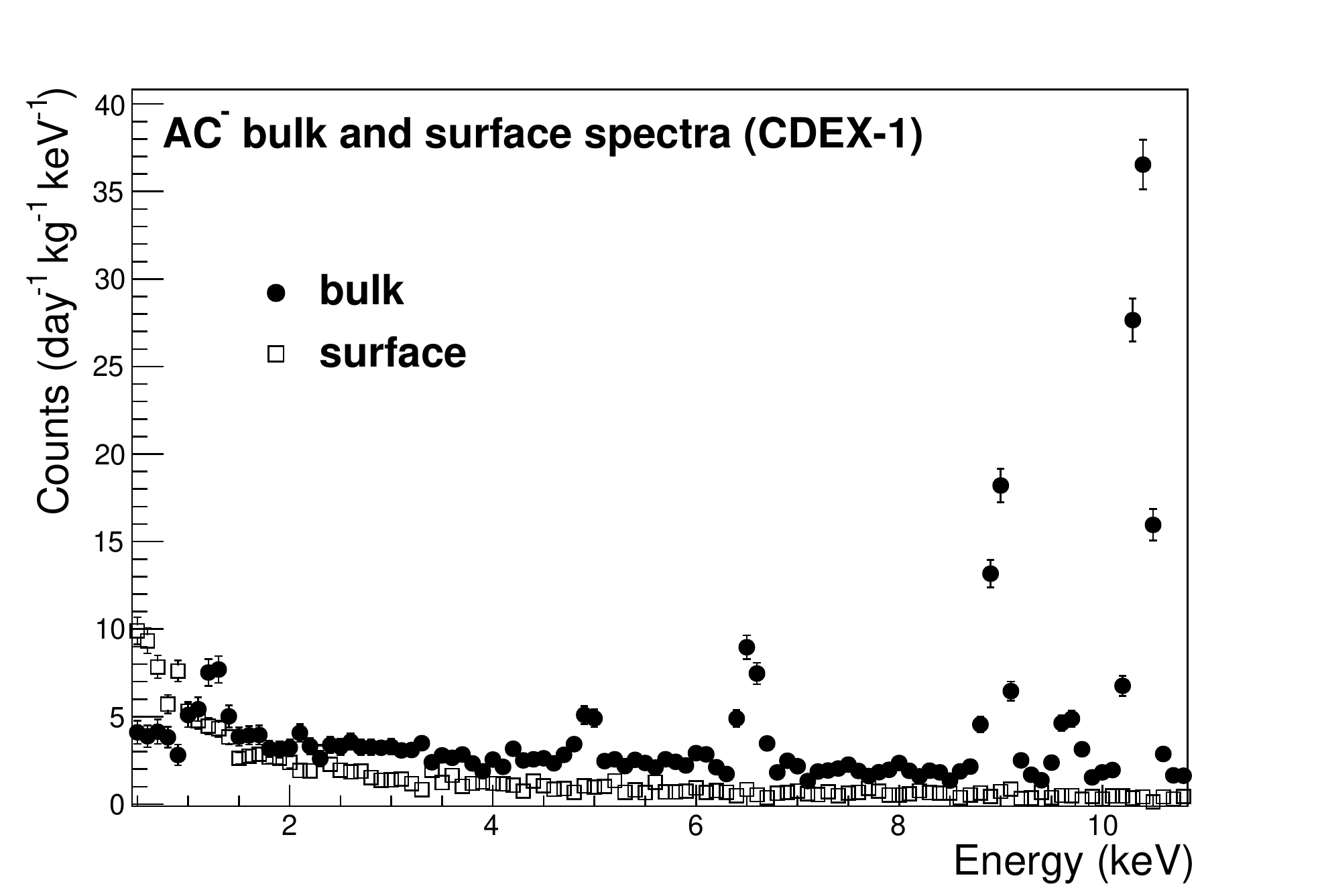}
  \caption{}
\end{subfigure}%
\newline
\begin{subfigure}{0.49\textwidth}
  \centering
  \includegraphics[width=1.1\linewidth]{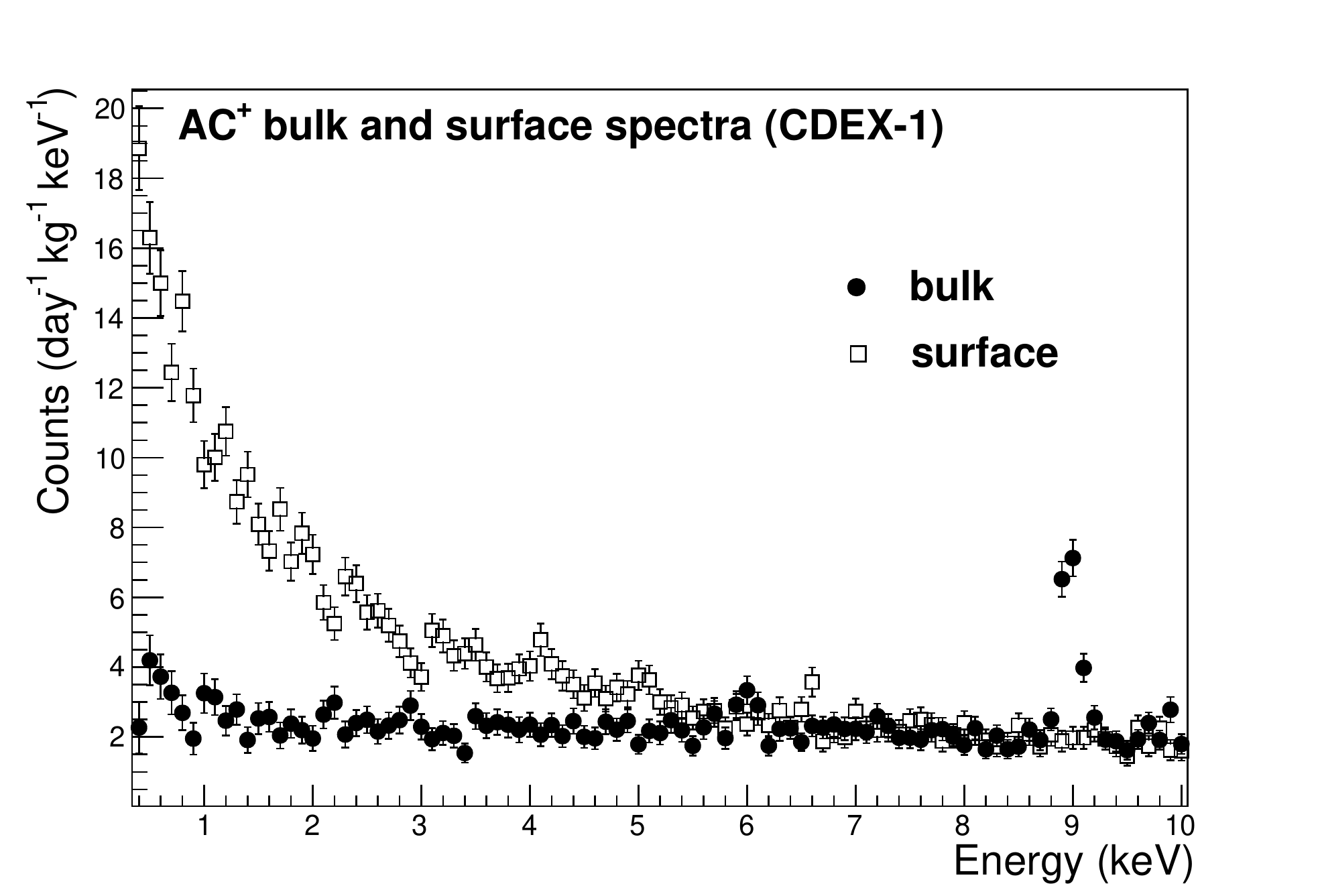}
  \caption{}
\end{subfigure}
\begin{subfigure}{0.49\textwidth}
  \centering
  \includegraphics[width=1.1\linewidth]{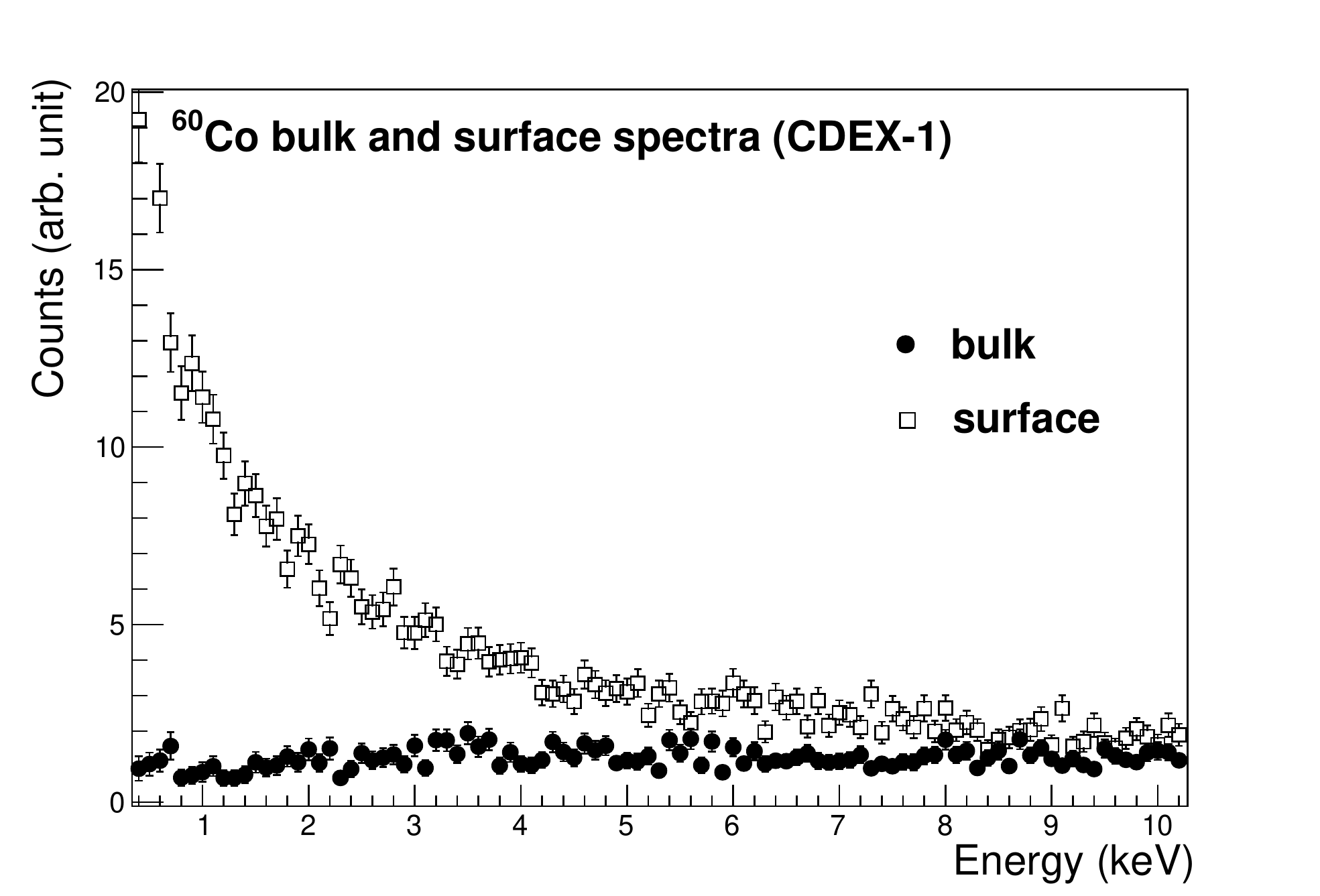}
  \caption{}
\end{subfigure}
\caption{
Bulk and surface spectra 
($B_{r i} (E)$ and $S_{r i} (E)$, respectively)
for (a) $\rm{AC}^{-}$, (b) $\rm{AC}^{+}$ 
         and (c) $^{60}\rm{Co}$. 
All X-ray peaks are correctly assigned to
bulk spectra, and all surface spectra show
monotonic increase at low energy, as expected.
}
\label{fig:ACV_ACT_Co60_Br_Sr}
\end{figure}

In addition to slower rise-time,
the S-events are also characterized by
incomplete charge collection, which manifests 
as spectra with monotonic increase at low energy,
as verified in  Figure~\ref{fig:ACV_ACT_Co60_Br_Sr} 
show that the $S_{r i} (E)$ spectra for
$i=\{ {\rm AC ^-} , {\rm AC ^+} , {\rm ^{60}Co} \}$.
In comparison, the $B_{r i} (E)$  
of the same channels are flat, as expected 
from their origins of Compton scattering.

\subsection{TEXONO data}
\label{section:texono_data}

\begin{figure}[!htbp]
\centering\includegraphics[width=0.8\linewidth]{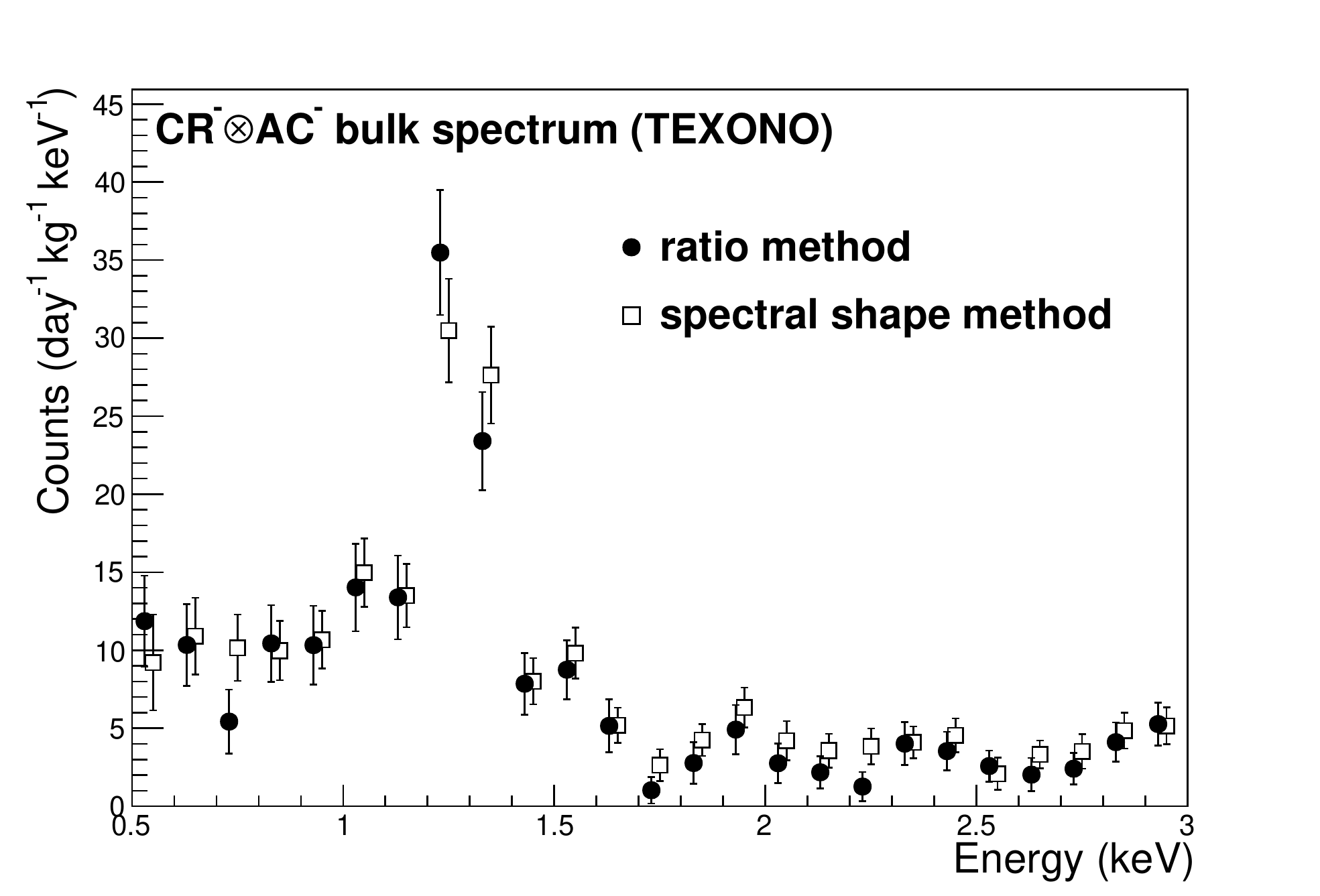}
\caption{
The $B_{r i} (E)$ spectra 
on the $\rm{CR}^{-}\otimes\rm{AC}^{-}$ channel from 
TEXONO data, with both spectral shape  and ratio methods.
The ratio method relies exclusively on
{\it in situ} data in this case.
}
\label{fig:k62_old_new_VrVBr}
\end{figure}

Re-analyses of 
published data from the TEXONO experiment~\cite{Li:2013a} 
were also analyzed with the ratio method.
Unlike the procedures for CDEX-1 analysis discussed in 
Section~\ref{section:cdex_1a_results}, 
external calibration sources are {\it not} used.
Instead, the analysis relies exclusive on 
all four categories of {\it in situ} event samples:
$\rm{CR}^{-}\otimes\rm{AC}^{-}$,
$\rm{CR}^{+}\otimes\rm{AC}^{-}$, $\rm{CR}^{-}\otimes\rm{AC}^{+}$ and
$\rm{CR}^{+}\otimes\rm{AC}^{+}$
which combine physics candidate samples as well as events 
due to ambient $\gamma$-rays and cosmic high energy neutrons.
As illustrated in Figure~\ref{fig:k62_old_new_VrVBr}, 
consistent results have been achieved with
the spectral shape method which
used $^{241}\rm{Am}$ and $^{137}\rm{Cs}$ sources 
as well as the
$\rm{CR}^{+}\otimes\rm{AC}^{-}$ data of 
n-type point-contact germanium detector~\cite{Li:2014a}.

%%new v24
It can be seen from Figure~\ref{fig:c1a_old_new_ACVBr} 
and \ref{fig:k62_old_new_VrVBr}
that both analysis methods on bulk-surface events identification with
TEXONO~\cite{Li:2013a} and CDEX-1~\cite{Yue:2014a,Zhao:2016}
data, respectively, give rise to consistent results.
%%old v24
%%The comparison shows that previous published results
%%of TEXONO~\cite{Li:2013a} and CDEX-1~\cite{Yue:2014a,Zhao:2016} 
%%are consistent with the earlier methods within error bars.

\section{Summary and Prospects}\label{section:discussion}

The ratio method provides an alternative way to address
the BSD problem in {\it p}Ge. 
Results consistent with the previous spectral shape method
are obtained, demonstrating its validity. 
Both methods are based on the assumption that the B- and S-events 
rise-time distributions (respectively, $f_B ( E , \tau )$
and $f_S ( E , \tau )$)
are similar among the adopted data samples. 

This feature is satisfied for B-events since nuclear- and electron-recoil 
events cannot be differentiated by their pulse shapes in germanium 
ionization detectors. The condition is also met for S-events at low 
near-threshold energy, the crucial energy range of interest, where the 
resolution smearing effects would dominate over intrinsic differences of 
their pulse shapes. At high energy, the requirement is matched with choice 
of calibration sources with consistent rise-time distributions as the 
physics samples. Systematic uncertainties as a result of such selection 
are evaluated, and then combined to derive the total uncertainties budget.

The most important merit of the ratio method is that the calibration 
can be achieved with {\it in situ} data, which can be neutrino- and 
WIMP-induced candidate events, cosmic-ray induced or ambient radioactivity 
background. This feature reduces or eliminates the dependence on 
external $\gamma$-ray sources for calibration purposes and therefore facilitates 
long-term data taking and operation of large multi-detector experiments. 
A drawback of the ratio method is the necessity to work in two-dimensional 
binning $( E , \tau )$, so that statistics have to be shared among many bins. 
The weight would be limited by the finite {\it in situ} counts in low background 
experiments. This gives rise to the choice for relatively large 
bin-size in 0.5 (${\rm{log_{10}({\mu}s)}}$) in the analysis.

Complete and accurate simulation of the Ge-detector behavior can provide 
complementary cross-checks to the calibration procedures and potentially 
improve on the systematic uncertainties. However, precise simulations 
of the rise-time distribution require 
many input parameters, some of which are not accurately known. In particular, 
the impurity levels and the leakage currents of the Ge-bulk crystal as well as 
details of resolution effects are crucial to the drift speed and hence the 
rise-time distributions. This is further complicated by possible time 
variations of the parameters. Simulation output at the current levels of 
sophistications do account for the qualitative behavior, but fall short of 
providing accurate quantitative descriptions of the measurement, as
compared to the use of {\it in situ} calibration data discussed in this work. 
Refining in the physics parameters input and advancing on the simulation
studies would be directions of future research.

\section{Acknowledgment}
This work is supported by the Academia Sinica Investigator Award 2011-15,
contracts 103-2112-M-001-024 and 104-2112-M-001-038-MY3 from the Ministry
of Science and Technology of Taiwan and the National Natural Science Foundation of
China (Nos.11175099, 11275107, 11475117, 11475099 and 11475092) and the National Basic
Research Program of China (973 Program) (2010CB833006) and the Tsinghua University
Initiative Scientific Research Program No.20121088494.

\bibliographystyle{model1-num-names}
\bibliography{new_bs_methods_v24}

\newcommand{\noop}[1]{}
\begin{thebibliography}{31}
\expandafter\ifx\csname natexlab\endcsname\relax\def\natexlab#1{#1}\fi
\providecommand{\bibinfo}[2]{#2}
\ifx\xfnm\relax \def\xfnm[#1]{\unskip,\space#1}\fi
%Type = Article
\bibitem[{Luke et~al.(1989)}]{Luke:1989}
\bibinfo{author}{P.~N. Luke}, et~al.,
\newblock \bibinfo{title}{Low capacitance large volume shaped-field germanium
  detector},
\newblock \bibinfo{journal}{IEEE Trans. Nucl. Sci.} \bibinfo{volume}{36}
  (\bibinfo{year}{1989}) \bibinfo{pages}{926}.
%Type = Article
\bibitem[{Barbeau et~al.(2007)}]{Barbeau:2007}
\bibinfo{author}{P.~S. Barbeau}, et~al.,
\newblock \bibinfo{title}{Large-mass ultralow noise germanium detectors:
  performance and applications in neutrino and astroparticle physics},
\newblock \bibinfo{journal}{J. Cosmol. Astropart. Phys.} \bibinfo{volume}{09}
  (\bibinfo{year}{2007}) \bibinfo{pages}{009}.
%Type = Article
\bibitem[{Liu et~al.(2017)}]{Liu:2017}
\bibinfo{author}{S.~K. Liu}, et~al.,
\newblock \bibinfo{title}{Constraints on {A}xion couplings from the {CDEX-1}
  experiment at the {C}hina {J}inping {U}nderground {L}aboratory},
\newblock \bibinfo{journal}{Phys. Rev. D} \bibinfo{volume}{95}
  (\bibinfo{year}{2017}) \bibinfo{pages}{052006}.
%Type = Article
\bibitem[{Yue et~al.(2004)}]{Yue:2004}
\bibinfo{author}{Q.~Yue}, et~al.,
\newblock \bibinfo{title}{Detection of {WIMPs} using low threshold {HPGe}
  detector},
\newblock \bibinfo{journal}{High Energy Phys. Nucl. Phys.} \bibinfo{volume}{28}
  (\bibinfo{year}{2004}) \bibinfo{pages}{877}.
%Type = Article
\bibitem[{Wong et~al.(2006)}]{Wong:2006}
\bibinfo{author}{H.~T. Wong}, et~al.,
\newblock \bibinfo{title}{Research program towards observation of
  neutrino-nucleus coherent scattering},
\newblock \bibinfo{journal}{J. Phys. Conf. Ser.} \bibinfo{volume}{39}
  (\bibinfo{year}{2006}) \bibinfo{pages}{266}.
%Type = Article
\bibitem[{Soma et~al.(2016)}]{Soma:2016}
\bibinfo{author}{A.~K. Soma}, et~al.,
\newblock \bibinfo{title}{Characterization and performance of germanium
  detectors with {sub-keV} sensitivities for neutrino and dark matter
  experiments},
\newblock \bibinfo{journal}{Nucl. Instr. Meth. Phys. Res. A}
  \bibinfo{volume}{836} (\bibinfo{year}{2016}) \bibinfo{pages}{67--82}.
%Type = Article
\bibitem[{Aalseth et~al.(2011)}]{Aalseth:2011a}
\bibinfo{author}{C.~E. Aalseth}, et~al.,
\newblock \bibinfo{title}{Results from a search for light-mass dark matter with
  a p-type point contact germanium detector},
\newblock \bibinfo{journal}{Phys. Rev. Lett.} \bibinfo{volume}{106}
  (\bibinfo{year}{2011}) \bibinfo{pages}{131301}.
%Type = Article
\bibitem[{Aalseth et~al.(2013)}]{Aalseth:2013}
\bibinfo{author}{C.~E. Aalseth}, et~al.,
\newblock \bibinfo{title}{{CoGeNT}: A search for low-mass dark matter using
  p-type point contact germanium detectors},
\newblock \bibinfo{journal}{Phys. Rev. D} \bibinfo{volume}{88}
  (\bibinfo{year}{2013}) \bibinfo{pages}{012002}.
%Type = Unpublished
\bibitem[{Aalseth et~al.(2014)}]{Aalseth:2014}
\bibinfo{author}{C.~E. Aalseth}, et~al., \bibinfo{title}{Search for an annual
  modulation in three years of {CoGeNT} dark matter detector data},
  \bibinfo{year}{2014}. \bibinfo{note}{ArXiv:1401.3295}.
%Type = Article
\bibitem[{Zhao et~al.(2013)}]{Zhao:2013}
\bibinfo{author}{W.~Zhao}, et~al.,
\newblock \bibinfo{title}{First results on low-mass {WIMPs} from the {CDEX-1}
  experiment at the {C}hina {J}inping {U}nderground {L}aboratory},
\newblock \bibinfo{journal}{Phy. Rev. D} \bibinfo{volume}{88}
  (\bibinfo{year}{2013}) \bibinfo{pages}{052004}.
%Type = Article
\bibitem[{Yue et~al.(2014)}]{Yue:2014a}
\bibinfo{author}{Q.~Yue}, et~al.,
\newblock \bibinfo{title}{Limits on light weakly interacting massive particles
  from the {CDEX-1} experiment with a p-type point-contact germanium detector
  at the {C}hina {J}inping {U}nderground {L}aboratory},
\newblock \bibinfo{journal}{Phys. Rev. D} \bibinfo{volume}{90}
  (\bibinfo{year}{2014}) \bibinfo{pages}{091701(R)}.
%Type = Article
\bibitem[{Zhao et~al.(2016)}]{Zhao:2016}
\bibinfo{author}{W.~Zhao}, et~al.,
\newblock \bibinfo{title}{Search of low-mass {WIMPs} with a p-type point
  contact germanium detector in the {CDEX-1} experiment},
\newblock \bibinfo{journal}{Phys. Rev. D} \bibinfo{volume}{93}
  (\bibinfo{year}{2016}) \bibinfo{pages}{092003}.
%Type = Article
\bibitem[{Li et~al.(2013)}]{Li:2013a}
\bibinfo{author}{H.~B. Li}, et~al.,
\newblock \bibinfo{title}{Limits on spin-independent couplings of {WIMP} dark
  matter with a p-type point-contact germanium detector},
\newblock \bibinfo{journal}{Phys. Rev. Lett.} \bibinfo{volume}{110}
  (\bibinfo{year}{2013}) \bibinfo{pages}{261301}.
%Type = Article
\bibitem[{Li et~al.(2014)}]{Li:2014a}
\bibinfo{author}{H.~B. Li}, et~al.,
\newblock \bibinfo{title}{Differentiation of bulk and surface events in p-type
  point-contact germanium detectors for light {WIMP} searches},
\newblock \bibinfo{journal}{Astropart. Phys.} \bibinfo{volume}{56}
  (\bibinfo{year}{2014}) \bibinfo{pages}{1--8}.
%Type = Article
\bibitem[{Martin et~al.(2012)}]{Martin:2012}
\bibinfo{author}{R.~D. Martin}, et~al.,
\newblock \bibinfo{title}{Determining the drift time of charge carriers in
  p-type point-contact {HPGe} detectors},
\newblock \bibinfo{journal}{Nucl. Instr. Meth. Phys. Res. A}
  \bibinfo{volume}{678} (\bibinfo{year}{2012}) \bibinfo{pages}{98--104}.
%Type = Article
\bibitem[{Aguayo et~al.(2013)}]{Aguayo:2013}
\bibinfo{author}{E.~Aguayo}, et~al.,
\newblock \bibinfo{title}{Characteristics of signals originating near the
  lithium-diffused {N+} contact of high purity germanium p-type point contact
  detectors},
\newblock \bibinfo{journal}{Nucl. Instr. Meth. Phys. Res. A}
  \bibinfo{volume}{701} (\bibinfo{year}{2013}) \bibinfo{pages}{176 -- 185}.
%Type = Unpublished
\bibitem[{Aalseth et~al.(2015)}]{Aalseth:2015a}
\bibinfo{author}{C.~E. Aalseth}, et~al., \bibinfo{title}{Maximum likelihood
  signal extraction method applied to 3.4 years of {CoGeNT} data},
  \bibinfo{year}{2015}. \bibinfo{note}{ArXiv:1401.6234v3}.
%Type = Article
\bibitem[{Jiang et~al.(2016)}]{Jiang:2016}
\bibinfo{author}{H.~Jiang}, et~al.,
\newblock \bibinfo{title}{Measurement of the dead layer thickness in a p-type
  point contact germanium detector},
\newblock \bibinfo{journal}{Chin. Phys. C} \bibinfo{volume}{40}
  (\bibinfo{year}{2016}) \bibinfo{pages}{096001}.
%Type = Article
\bibitem[{Ma et~al.(2017)}]{Ma:2017}
\bibinfo{author}{J.~L. Ma}, et~al.,
\newblock \bibinfo{title}{Study of inactive layer uniformity and charge
  collection efficiency of a p-type point-contact germanium detector},
\newblock \bibinfo{journal}{Appl. Radiat. Isot.} \bibinfo{volume}{127}
  (\bibinfo{year}{2017}) \bibinfo{pages}{130--136}.
%Type = Article
\bibitem[{Baudis et~al.(1998)}]{Baudis:1998}
\bibinfo{author}{L.~Baudis}, et~al.,
\newblock \bibinfo{title}{High-purity germanium detector ionization pulse
  shapes of nuclear recoils, $\gamma$-interactions and microphonism},
\newblock \bibinfo{journal}{Nucl. Instr. Meth. Phys. Res. A}
  \bibinfo{volume}{418} (\bibinfo{year}{1998}) \bibinfo{pages}{348—354}.
%Type = Article
\bibitem[{Wei et~al.(2016)}]{Wei:2016}
\bibinfo{author}{W.~Z. Wei}, et~al.,
\newblock \bibinfo{title}{Discrimination of nuclear and electronic recoil
  events using plasma effect in germanium detectors},
\newblock \bibinfo{journal}{J. Inst.} \bibinfo{volume}{11}
  (\bibinfo{year}{2016}) \bibinfo{pages}{P07008}.
%Type = Article
\bibitem[{Singh et~al.(2017)}]{Singh:2017}
\bibinfo{author}{M.~K. Singh}, et~al.,
\newblock \bibinfo{title}{Design and performance of a hybrid fast and thermal
  neutron detector},
\newblock \bibinfo{journal}{Nucl. Instr. Meth. Phys. Res. A}
  \bibinfo{volume}{868} (\bibinfo{year}{2017}) \bibinfo{pages}{109--118}.
%Type = Misc
\bibitem[{Sonay(shed)}]{Sonay:2017}
\bibinfo{author}{A.~Sonay}, \bibinfo{title}{Characterization of neutron and
  high purity germanium detectors with advanced data acquisition system and
  measurement of neutron background at the {K}uo-{S}heng neutrino laboratory,
  {M}. {S}c. thesis, {D}okuz {E}yl\"{u}l {U}niversity, {T}urkey},
  \bibinfo{year}{2018; to be published}.
%Type = Article
\bibitem[{Agnese et~al.(2016)}]{Agnese:2016}
\bibinfo{author}{R.~Agnese}, et~al.,
\newblock \bibinfo{title}{New results from the search for low-mass weakly
  interacting massive particles with the {CDMS} low ionization threshold
  experiment},
\newblock \bibinfo{journal}{Phys. Rev. Lett.} \bibinfo{volume}{116}
  (\bibinfo{year}{2016}) \bibinfo{pages}{071301}.
%Type = Article
\bibitem[{Angloher et~al.(2016)}]{Angloher:2016}
\bibinfo{author}{G.~Angloher}, et~al.,
\newblock \bibinfo{title}{Results on light dark matter particles with a
  low-threshold {CRESST-II} detector},
\newblock \bibinfo{journal}{Eur. Phys. J. C} \bibinfo{volume}{76}
  (\bibinfo{year}{2016}) \bibinfo{pages}{25}.
%Type = Article
\bibitem[{Agnese et~al.(2014)}]{Agnese:2014}
\bibinfo{author}{R.~Agnese}, et~al.,
\newblock \bibinfo{title}{Search for low-mass weakly interacting massive
  particles with {SuperCDMS}},
\newblock \bibinfo{journal}{Phys. Rev. Lett.} \bibinfo{volume}{112}
  (\bibinfo{year}{2014}) \bibinfo{pages}{241302}.
%Type = Article
\bibitem[{Tan et~al.(2016)}]{Tan:2016}
\bibinfo{author}{A.~Tan}, et~al.,
\newblock \bibinfo{title}{Dark matter results from first 98.7 days of data from
  the {P}anda{X}-{II} experiment},
\newblock \bibinfo{journal}{Phys. Rev. Lett.} \bibinfo{volume}{117}
  (\bibinfo{year}{2016}) \bibinfo{pages}{121303}.
%Type = Article
\bibitem[{Akerib et~al.(2017)}]{Akerib:2017}
\bibinfo{author}{D.~S. Akerib}, et~al.,
\newblock \bibinfo{title}{Results from a search for dark matter in the complete
  {LUX} exposure},
\newblock \bibinfo{journal}{Phys. Rev. Lett.} \bibinfo{volume}{118}
  (\bibinfo{year}{2017}) \bibinfo{pages}{021303}.
%Type = Article
\bibitem[{Belli et~al.(2011)}]{Belli:2011}
\bibinfo{author}{P.~Belli}, et~al.,
\newblock \bibinfo{title}{Observations of annual modulation in direct detection
  of relic particles and light neutralinos},
\newblock \bibinfo{journal}{Phys. Rev. D} \bibinfo{volume}{84}
  (\bibinfo{year}{2011}) \bibinfo{pages}{055014}.
%Type = Article
\bibitem[{Bernabei et~al.(2010)}]{Bernabei:2010}
\bibinfo{author}{R.~Bernabei}, et~al.,
\newblock \bibinfo{title}{New results from {DAMA}/{LIBRA}},
\newblock \bibinfo{journal}{Eur. Phys. J. C} \bibinfo{volume}{67}
  (\bibinfo{year}{2010}) \bibinfo{pages}{39}.
%Type = Article
\bibitem[{Agnese et~al.(2013)}]{Agnese:2013}
\bibinfo{author}{R.~Agnese}, et~al.,
\newblock \bibinfo{title}{Silicon detector dark matter results from the final
  exposure of {CDMS II}},
\newblock \bibinfo{journal}{Phys. Rev. Lett.} \bibinfo{volume}{111}
  (\bibinfo{year}{2013}) \bibinfo{pages}{251301}.

\end{thebibliography}

\end{document}